\documentclass[aps,prl,twocolumn,superscriptaddress]{revtex4-2}

\usepackage{graphicx}
\usepackage{float}
\usepackage[colorlinks=true,linkcolor=blue,citecolor=red]{hyperref}
\usepackage{siunitx}

\usepackage[justification=justified]{caption}
\usepackage{ragged2e}
\usepackage{braket}
\usepackage{chemformula}

\DeclareCaptionLabelFormat{mylabel}{#1#2.\hspace{1.5ex}}
\usepackage[nameinlink]{cleveref}
\Crefname{equation}{Eq.}{Eqs.}
\Crefname{figure}{Fig.}{Figs.}

\usepackage{array}    

\usepackage{xr}
\makeatletter

\newcommand*{\addFileDependency}[1]{
\typeout{(#1)}
\@addtofilelist{#1}
\IfFileExists{#1}{}{\typeout{No file #1.}}
}\makeatother

\begin{document}

\title{Quantum Noise Spectroscopy of Nanoscale Charge Defects in Silicon Carbide at Room Temperature}

\author{Jinpeng Liu}%
\email{These authors contributed equally to this work.}
\affiliation{School of Biomedical
Engineering, Division of Life Sciences and Medicine, University of Science and
Technology of China, Hefei, Anhui, 230026, P.R.China}
\affiliation{Laboratory of Spin Magnetic Resonance, School of Physical Sciences, Anhui Province Key Laboratory of Scientific Instrument Development and Application, University of Science and Technology of China, Hefei 230026, China}
\affiliation{Suzhou Institute for
Advanced Research, University of Science and Technology of China, Suzhou,
Jiangsu, 215123, P.R.China}
\author{Yuanhong Teng}
\email{These authors contributed equally to this work.}
\affiliation{Laboratory of Spin Magnetic Resonance, School of Physical Sciences, Anhui Province Key Laboratory of Scientific Instrument Development and Application, University of Science and Technology of China, Hefei 230026, China}

\author{Yu Chen}
\email{These authors contributed equally to this work.}
\affiliation{Laboratory of Spin Magnetic Resonance, School of Physical Sciences, Anhui Province Key Laboratory of Scientific Instrument Development and Application, University of Science and Technology of China, Hefei 230026, China}
\affiliation{Hefei National Laboratory, University of Science and Technology of China, Hefei 230088, China}

\author{Yixuan Wang}
 \affiliation{Laboratory of Spin Magnetic Resonance, School of Physical Sciences, Anhui Province Key Laboratory of Scientific Instrument Development and Application, University of Science and Technology of China, Hefei 230026, China}
\affiliation{Hefei National Laboratory, University of Science and Technology of China, Hefei 230088, China}

\author{Sen Luan}
\affiliation{School of Biomedical
Engineering, Division of Life Sciences and Medicine, University of Science and
Technology of China, Hefei, Anhui, 230026, P.R.China}
\affiliation{Laboratory of Spin Magnetic Resonance, School of Physical Sciences, Anhui Province Key Laboratory of Scientific Instrument Development and Application, University of Science and Technology of China, Hefei 230026, China}
\affiliation{Suzhou Institute for
Advanced Research, University of Science and Technology of China, Suzhou,
Jiangsu, 215123, P.R.China}

\author{Chihang Luo}
\affiliation{Laboratory of Spin Magnetic Resonance, School of Physical Sciences, Anhui Province Key Laboratory of Scientific Instrument Development and Application, University of Science and Technology of China, Hefei 230026, China}

\author{Jun Yin}%
\affiliation{School of Biomedical
Engineering, Division of Life Sciences and Medicine, University of Science and
Technology of China, Hefei, Anhui, 230026, P.R.China}
\affiliation{Laboratory of Spin Magnetic Resonance, School of Physical Sciences, Anhui Province Key Laboratory of Scientific Instrument Development and Application, University of Science and Technology of China, Hefei 230026, China}
\affiliation{Suzhou Institute for
Advanced Research, University of Science and Technology of China, Suzhou,
Jiangsu, 215123, P.R.China}

\author{Hao Li}
\affiliation{Shanghai Key Laboratory of Superconductor Integrated Circuit Technology, Shanghai Institute of Microsystem and Information Technology, Chinese Academy of Sciences, Shanghai 200050, China}

\author{Lixing You}
\affiliation{Shanghai Key Laboratory of Superconductor Integrated Circuit Technology, Shanghai Institute of Microsystem and Information Technology, Chinese Academy of Sciences, Shanghai 200050, China}
\author{Qiang Li}
\affiliation{State Key Laboratory of Silicon Materials and Advanced Semiconductors $\&$ School of Materials Science and Engineering, Institute of Advanced Semiconductors, ZJU-Hangzhou Global Scientific and Technological Innovation Center, Zhejiang University, Hangzhou，Zhejiang, 310027, China}
\author{Chunming Yin}

\author{Ya Wang}
\affiliation{Laboratory of Spin Magnetic Resonance, School of Physical Sciences, Anhui Province Key Laboratory of Scientific Instrument Development and Application, University of Science and Technology of China, Hefei 230026, China}

\affiliation{Hefei National Laboratory, University of Science and Technology of China, Hefei 230088, China}

\affiliation{Hefei National Research Center for Physical Sciences at the Microscale, University of Science and Technology of China, Hefei 230026, China}

\author{Qi Zhang}
\email{zhq2011@ustc.edu.cn}
\affiliation{School of Biomedical
Engineering, Division of Life Sciences and Medicine, University of Science and
Technology of China, Hefei, Anhui, 230026, P.R.China}

\affiliation{Suzhou Institute for
Advanced Research, University of Science and Technology of China, Suzhou,
Jiangsu, 215123, P.R.China}

\affiliation{Institute of Quantum Sensing, School of Physics, Institute of Fundamental and Transdisciplinary Research, Zhejiang Key Laboratory of R$\&$D and Application of Cutting-edge Scientific Instruments, Zhejiang University, Hangzhou, 310027, China}

\author{Fazhan Shi}
\email{fzshi@ustc.edu.cn}
\affiliation{School of Biomedical
Engineering, Division of Life Sciences and Medicine, University of Science and
Technology of China, Hefei, Anhui, 230026, P.R.China}

\affiliation{Laboratory of Spin Magnetic Resonance, School of Physical Sciences, Anhui Province Key Laboratory of Scientific Instrument Development and Application, University of Science and Technology of China, Hefei 230026, China}
\affiliation{Suzhou Institute for
Advanced Research, University of Science and Technology of China, Suzhou,
Jiangsu, 215123, P.R.China}
\affiliation{Hefei National Laboratory, University of Science and Technology of China, Hefei 230088, China}

\affiliation{Hefei National Research Center for Physical Sciences at the Microscale, University of Science and Technology of China, Hefei 230026, China}

\begin{abstract}
The nanoscale charge environment critically influences the performance of semiconductor devices, yet traditional characterization techniques usually lack the spatial resolution to resolve nanoscale charge heterogeneity and identify microscopic noise sources. Here, we address this critical gap by utilizing individual PL5 color centers in commercial silicon carbide as room-temperature, broadband (near-DC to GHz) quantum spectrometers. We report the first real-time, nanoscale observation of single-charge tunneling dynamics at room temperature in 4H-SiC by monitoring the random telegraph noise using optically detected magnetic resonance (ODMR). This capability enables an electrical noise imaging technique that shows distinct noise variations on different wafers. Employing dynamical decoupling and $T_1$ relaxation spectroscopy, we observe correlations of noise spectral density across frequency bands and obtain the nanoscale electron paramagnetic resonance (EPR) fingerprint of spin defects in SiC. Ultimately, through surface chemical engineering, we identify the native oxide layer as the main source of resolved electromagnetic fluctuations. This work establishes a versatile atomic-scale metrology platform bridging quantum sensing and semiconductor engineering, offering critical insights for optimizing third-generation semiconductor interfaces and advancing solid-state quantum technologies.
\end{abstract}


\maketitle

\section{Introduction}

Silicon carbide is recognized as a promising semiconductor material due to its outstanding physical properties, such as high thermal conductivity and wide band gap, making it ideal for the fabrication of high-power devices \cite{2017-ReviewSiliconCarbide-She-IEEETrans.Ind.Electron.}. However, the performance and reliability of SiC-based complementary metal-oxide-semiconductor (CMOS) devices are fundamentally limited by atomic-scale defects in bulk and at the interface, which act as noise sources, reduce carrier lifetimes, and trigger breakdown \cite{2022-DefectInspectionTechniques-Chen-NanoscaleResLett}. Despite the maturity of SiC fabrication, characterizing the microscopic electric and magnetic environments associated with these defects presents a critical challenge. Conventional methods like standard deep level transient spectroscopy (DLTS), electron spin resonance (ESR) and macroscopic photoluminescence (PL) typically provide volume-averaged metrics  \cite{2022-CharacterizationMethodsDefects-Bathen-J.Appl.Phys.b}, inherently preventing the identification and localization of noise sources at the nanoscale.

\begin{figure*} [htbp]
\includegraphics{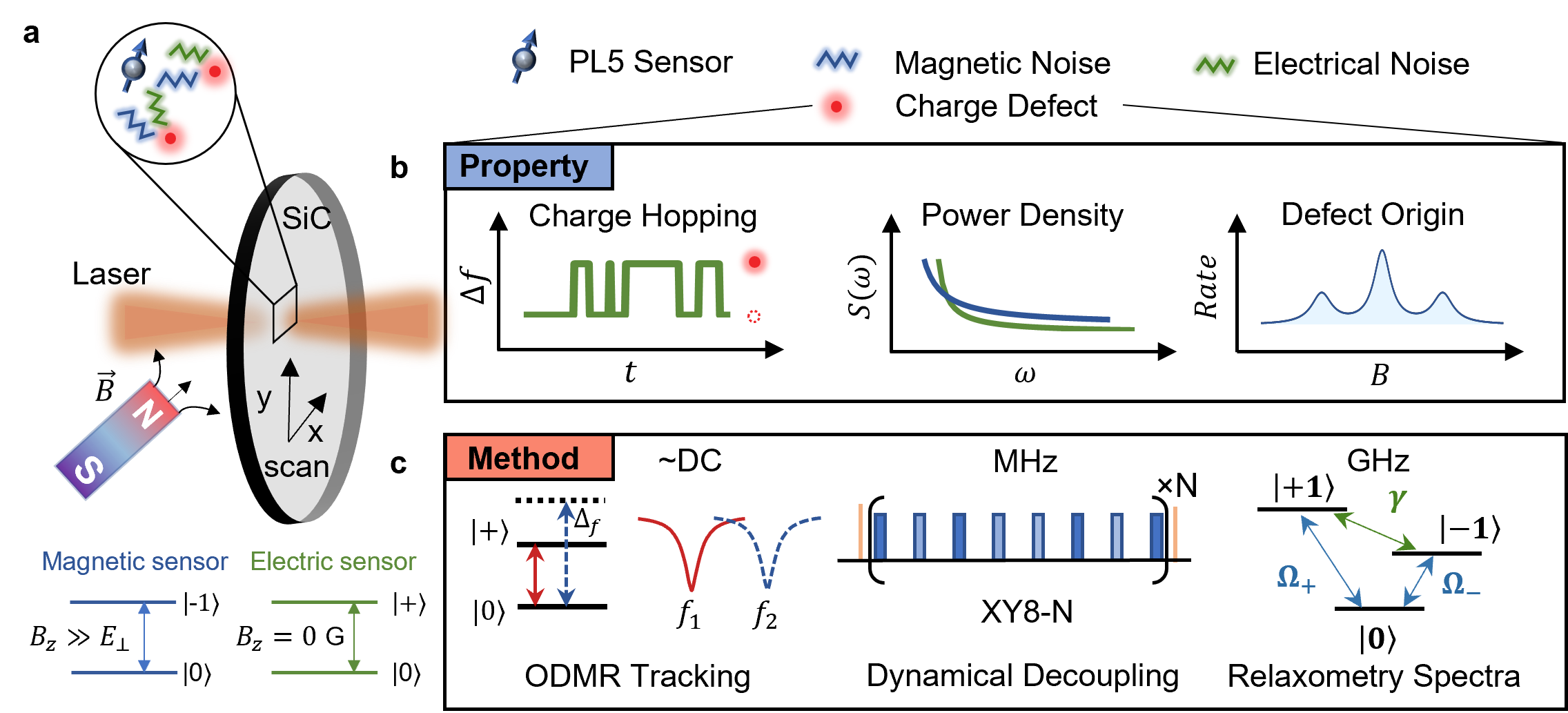}
\caption{\justifying \textbf{Experimental setup and sensing scheme overview.} (a) Experiment principle of quantum sensor in silicon carbide. A 905nm laser is used to realize the initialization and readout of PL5 center. Electrical and magnetic noise in semiconductors can have a significant impact on the performance of high-power electronic devices as well as on the coherence properties of qubits embedded in solid-state materials. An external magnetic field parallel to the principal axis of PL5 center is applied to tune the energy levels when needed. The intrinsic none-zero term $E_x$ for PL5 makes the transition between ground states $\{\ket{0},\ket{+}\}$ exclusively sensitive to transverse electric noise under zero magnetic field. When $B_z$ is applied, the transition become sensitive to longitudinal magnetic noise. (b) Physical properties of the local charge environment. The panels illustrate three key dimensions of defect characterization: (Left) Charge hopping, showing the time-domain random telegraph noise induced by discrete charge trapping and releasing events; (Middle) Power Density, displaying the frequency-domain noise spectrum 
for both magnetic (blue) and electrical (green) noise components; and (Right) Defect Species, revealing the spectroscopic fingerprint used to identify specific defects. (c) Corresponding quantum sensing protocols. The methods span from DC to GHz frequencies: (Left) Real-time tracking of ODMR resonance shifts $\Delta f$ monitor low-frequency fluctuations; (Middle) Dynamical decoupling sequences (e.g., XY8) with variable inter-pulse spacing to probe noise power density in the MHz regime; and (Right) 
$T_1$ relaxometry-based EPR spectroscopy to probe charge defects species in the GHz regime.}
\label{fig1}
\end{figure*}
To fill this gap, solid-state quantum sensors have emerged as promising tools for nanoscale diagnostics. While nitrogen-vacancy (NV) centers in diamond are widely employed for nanoscale electric field sensing \cite{2014-NanoscaleDetectionSingle-Dolde-Phys.Rev.Lett.,2011-ElectricfieldSensingUsing-Dolde-NaturePhys,2016-OpticalPatterningTrapped-Jayakumar-NatCommun,2021-OpticalActivationDetection-Lozovoi-NatElectrona,2024-3DMappingManipulationPhotocurrent-Wood-AdvancedMaterials,2024-CorrelatedSpectroscopyElectric-Delord-NanoLett.,2024-CorrelatedSensingSolidstate-Ji-Nat.Photon.,2024-TrackingSingleCharge-Weng-Phys.Rev.B,2022-ImagingDarkCharge-Lozovoi-Sci.Adv.,monge2025ensembleaveragingparallelizedsingleshot}, their application to SiC is constrained by the sensor-sample distance and the inability to probe the intrinsic bulk and interface environment of the semiconductor \cite{2024-SinglemoleculeScaleMagnetic-Du-Rev.Mod.Phys.,2020-SolutionElectricField-Oberg-Phys.Rev.Applied,RevModPhys.89.035002}. Moreover, single-charge tunneling events detected via photoluminescence excitation predominantly required cryogenic temperatures, thereby limiting their applicability to in-fab-line monitoring \cite{2019-ElectricalOpticalControl-Anderson-Science,2024-AtomicOpticalAntennas-Li-Nat.Photon.,2024-CorrelatedSensingSolidstate-Ji-Nat.Photon.,2025-QuantumElectrometerTimeresolved-Pieplow-NatureCommunicationsa}. To achieve true in-situ monitoring, we focus on utilizing intrinsic point defects within the 4H-SiC lattice itself. Among these, the PL5 center stands out not only for its optical addressability and spin coherence \cite{2011-RoomTemperatureCoherent-Koehl-Nature, 2024-RoomtemperatureWaveguideIntegrated-Hu-NatCommun,2022-RoomtemperatureCoherentManipulation-Li-Natl.Sci.Rev.,chen2026oxygenvacancyquantumspindefects} but, more importantly, for its giant Stark effect. The Stark-coupling parameters of PL5 center are 2–7 times larger than those of NV centers in diamond \cite{2014-ElectricallyMechanicallyTunable-Falk-Phys.Rev.Lett.}, making it an ultrasensitive electrometer capable of detecting local electric field fluctuations. This unique property makes PL5 center an ideal candidate for imaging the noise landscape inside SiC devices at room temperature.

Concurrently, recent research on solid-state spin qubits has revealed that, in addition to the magnetic spin bath, charge defects constitute a primary source of decoherence \cite{2019-ElectricalOpticalControl-Anderson-Science,zeledon2025minutelongquantumcoherenceenabled}. While existing studies have largely focused on monitoring and spatially localizing charge fluctuations in the DC-MHz regime at cryogenic temperatures \cite{2025-QuantumElectrometerTimeresolved-Pieplow-NatureCommunicationsa}, or suppressing noise via electric field depletion control \cite{2019-ElectricalOpticalControl-Anderson-Science,2025-SingleV2Defect-Steidl-NatCommun,zeledon2025minutelongquantumcoherenceenabled}, a method for broadband and species-resolved identification of nanoscale noise sources remains unrealized. This gap hinders the optimization of material growth and device fabrication processes for both the semiconductor industry and quantum technologies. Here, we bridge this gap by developing a room-temperature, broadband (DC to GHz) quantum noise spectrometer using individual PL5 centers in commercial 4H-SiC. Empowered by the extreme electric field sensitivity and nanoscale resolution, we capture real-time charge dynamics. Expanding from this microscopic insight to a macroscopic metrology tool, we achieve frequency-resolved characterization of the noise environment — from mapping the heterogeneities and quality between different wafers, to obtaining the nanoscale EPR fingerprint of spin defects hosted in the material. Finally, by comparing experimental results before and after surface treatment, we identify the main physical origin of the noise as the native SiO$_2$ layer on the surface.

\section{Results}

\begin{figure} [htbp]
\includegraphics{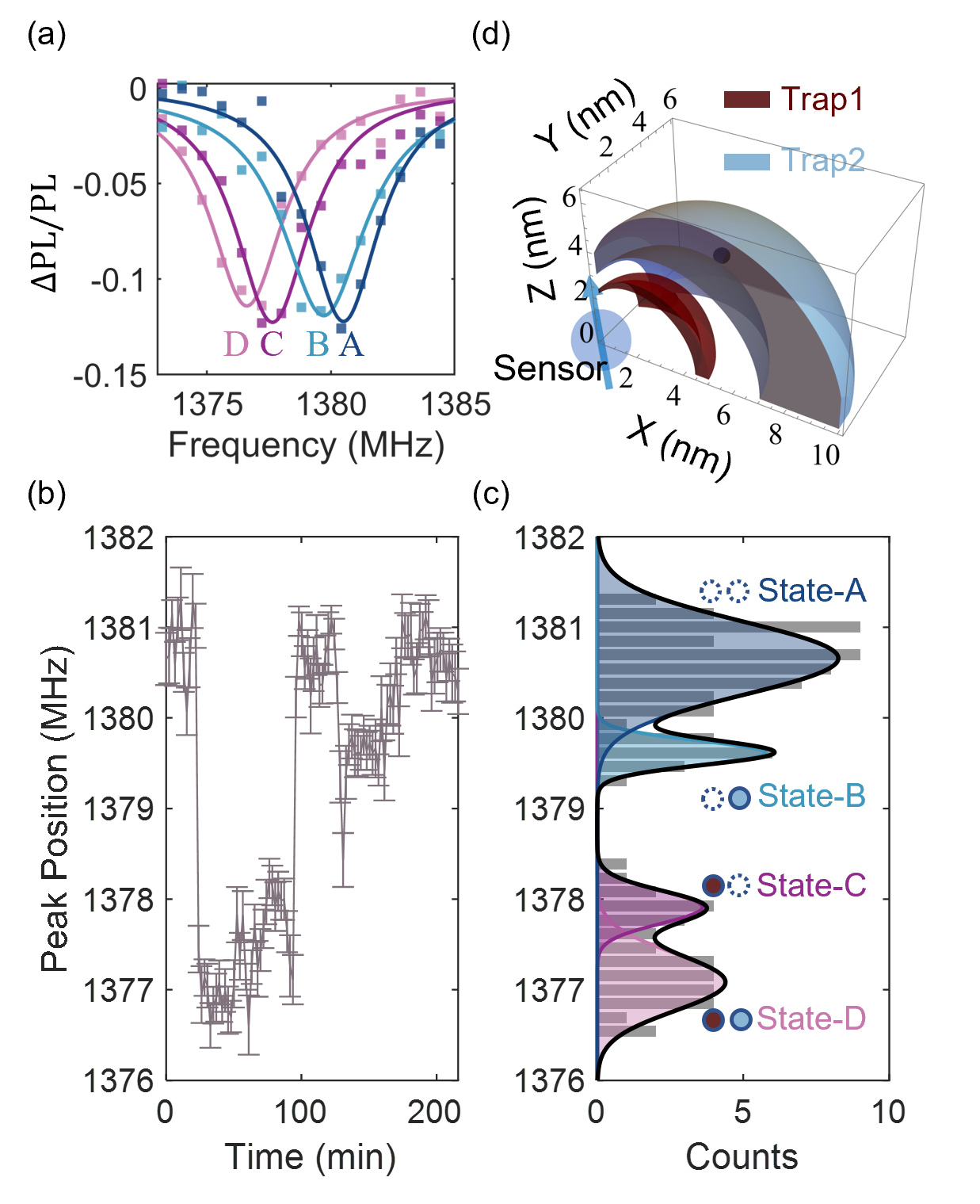}
\caption{\justifying \textbf{Nanoscale monitoring of electric environment at single
charge level}. (a) Four typical CW spectra configurations with differ-
ent central frequency of a single quantum sensor were observed, which
were labeled as A-D. (b) Time-resolved CW-ODMR measurements on
a single PL5 center reveal the evolution of its spectral peak position.
(c) Histogram illustrates the distribution of peak positions in (b). Four well-defined stable states can be clearly identified by fitting, corresponding to the four states identified in panel (a). The peak separation between State-A and C (B and D) is 3 $\pm$ 0.3 MHz, while that between State-A and B (C and D)is 1 $\pm$ 0.3 MHz. This indicates that the observed spectral fluctuations result from a composition of two two-level system (TLS) signals. (d) The possible location of the two charge traps with 95 \% confidence interval based on the splitting of different states shown in (c). The detail of the calculation can be found in Supplementary Material Section S2.3 \cite{SM}} 
\label{fig2}
\end{figure}

\textbf{Experimental setup and Sensing Scheme} A full landscape of our broadband noise detection technique routine in silicon carbide at room temperature is shown in \autoref{fig1}. Our sample was diced from a wafer consisting of an intrinsic epitaxial layer of single-crystal 4H-SiC grown on a $4^\circ$ off-axis N-type 4H-SiC substrate. To generate PL5 centers, 60-keV $^{14}\text{N}^{+}$ ions were implanted at a dose of $10^{10} \ \text{cm}^{-2}$ followed by annealing for 30 minutes. More details are provided in Supplementary Material Section S1.3 \cite{SM}. A 905 nm laser is used to initialize and read out the spin state of a single PL5 quantum sensor in 4H-SiC, where a fluorescence count rate of 250 kcps (kilo counts per second) for a single PL5 center has been observed without photonic structure enhancement. The Hamiltonian of PL5 center is given by 

\begin{equation}
H / h =\frac{1}{\hbar^2} \left( 
D \hat{S}_z^2
-  E_x(S_x^2-S_y^2)+\gamma\textbf{B}\cdot\textbf{S}
\right)
\end{equation} 
where $D\approx 1360$ MHz and $E_x\approx 16$ MHz (the coordinate is chosen to parallel the c-axis to the xz plane) \cite{chen2026oxygenvacancyquantumspindefects,2022-RoomtemperatureCoherentManipulation-Li-Natl.Sci.Rev.}. The Hamiltonian dictates that under zero magnetic field, the none-zero term $E_x$ makes the transition between ground states exclusively sensitive to transverse electric noise. Microwave pulses are applied to coherently manipulate the spin. An external magnetic field parallel to the principal axis of PL5 center (labeled as $B_z$) is applied to tune the energy levels when needed. As illustrated in \autoref{fig1}(b-c), we employ a suite of coherent control sequences—ranging from continuous-wave (CW) monitoring to dynamical decoupling and $T_1$ relaxometry—to filter and track spectral fluctuations across a wide frequency bandwidth (near-DC to GHz). This setup effectively converts the PL5 center into a localized broadband spectrum analyzer for the surrounding electrical and magnetic environment induced by charge defects .

\begin{figure*}[htbp]
\includegraphics{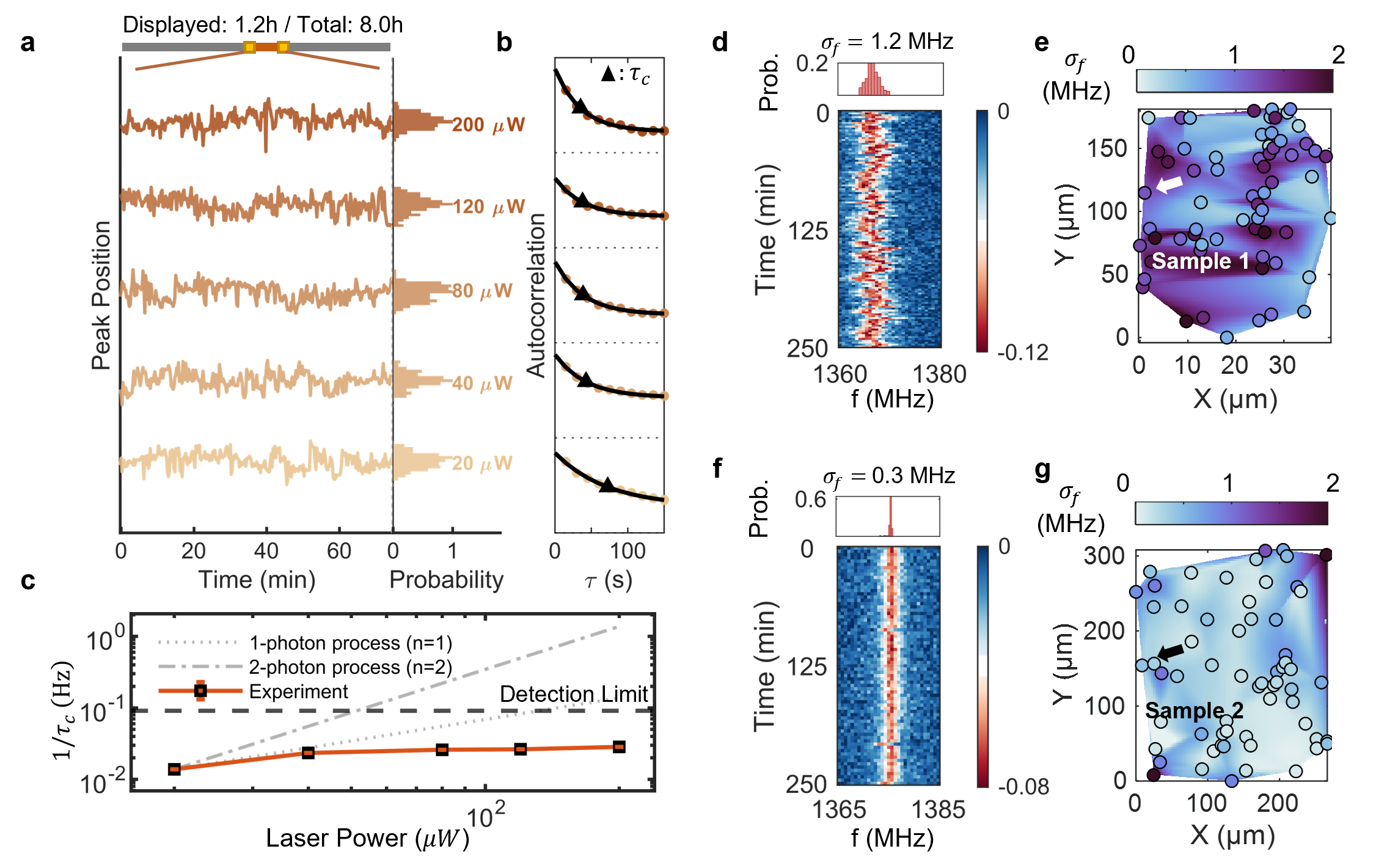}
\caption{\justifying \label{map}\textbf{Continuous charge jitter and nanoscale electrical noise imaging of commercial SiC wafers.} (a) Long-term continuous-wave ODMR tracking of a representative PL5 sensor under different laser powers. The sensor shows a random, continuous spectral wandering, which represents the typical noise behavior for most investigated defects. For visual clarity, only a portion of the data and its distribution at each laser power is displayed. (b) Autocorrelation function (ACF) of the spectral trajectories. Solid black lines are exponential fits used to extract the characteristic fluctuation time constant ($\tau_c$). Details of the ACF extraction and analysis are provided in the Supplementary Material. (c) Extracted fluctuation rate ($1/\tau_c$) as a function of laser power. Even as the optical power spans an order of magnitude, the fluctuation rate exhibits only a marginal increase, remaining consistently far below the theoretical rates expected for single- or two-photon processes. This result indicates that this continuous noise is driven by a thermally-activated background bath rather than direct optical excitation. The fit error bars ($1\sigma$) are smaller than the size of the data markers. (d) Spectral diffusion tracking and the corresponding probability distribution of a typical sensor from Wafer 1 (Sample 1), showing a broad peak distribution ($\sigma_f = 1.2$ MHz). (e) Nanoscale electrical noise map of Wafer 1. The spatial distribution of $\sigma_f$ reveals clear variations in the local charge environment over tens of microns. Colored maps are interpolated from the measured ($\sigma_f$) values at individual PL5 locations; circles mark the measured sensors. (f, g) Corresponding spectral tracking and noise map for Wafer 2 (Sample 2). Despite having the same fabrication history, Wafer 2 shows a much higher spectral stability ($\sigma_f = 0.3$ MHz). This contrast demonstrates the use of our quantum probe to evaluate intrinsic variations in wafer quality.}
\end{figure*}

\textbf{Electrometry of Two Charge Traps.} The change of local electric field will cause Stark shift between $\ket{0}$ and $\ket{+}/\ket{-}$ transitions as displayed in \autoref{fig1}(c), where $\ket{+}/\ket{-}=\frac{1}{\sqrt{2}}(\ket{+1}\pm \ket{-1})$. Due to the orientation of the microwave field, spin manipulation between the $\ket{0}$ and $\ket{+}$ is more effective under zero magnetic field in our setup.  For certain PL5 centers, we observed time-dependent spectral
switching in their CW-ODMR results as shown in \Cref{fig2}.
\Cref{fig2}(a) presents four representative ODMR spectral
configurations (labeled State A to State D) acquired from
the same individual color center. The raw data and their
corresponding Lorentzian fits clearly show that the sensor exhibits distinct resonance frequencies with shifted central frequencies at different moments, indicating a dynamic local environment. To systematically capture these dynamics, we performed a continuous spectral tracking measurement over a duration of several hours. The resulting
time-resolved map, shown in \Cref{fig2}(b), visualizes the
temporal evolution of the peak positions. By compiling
the resonance frequencies extracted from this long-term
trace, we obtained the statistical distribution presented
in \Cref{fig2}(c). We attribute this behavior to the surrounding traps, which cause fluctuation of electric field when they capture or release charges. Based on the spectral splitting between different states (3 $\pm$ 0.3 MHz between state A and C, 1 $\pm$ 0.3 MHz between state C and D), we constrained the possible locations of the two charge traps with 95 \% confidence interval using the Stark shift coeffcients of PL5 centers. Given the significantly larger transverse Stark coeffcient compared to its longitudinal part ($d_\perp$ = 32.5, $d_\parallel$  < 3 (Hz cm/V)) \cite{2014-ElectricallyMechanicallyTunable-Falk-Phys.Rev.Lett.}, we infer that the
observed electric field fluctuation originates from charge
trapping and releasing events of two traps within a region
of less than 10 nm [\Cref{fig2}(d)]. More analysis is shown in Supplementary Material Section S2 \cite{SM}.

\textbf{Noise Imaging and Material Characterization}
 While the charge hopping events observed in \autoref{fig2} provide a microscopic picture of single charge defects dynamics, such discrete jumping behavior is quite rare, appearing in less than $1\%$ of the hundreds of PL5 sensors we investigated. For the vast majority of sensors, the spectral instability appeared as quasi-continuous jitter where individual charge-transition events are hard to be clearly resolved. To systematically quantify this noise, we performed long-term CW-ODMR tracking on these representative unstable sensors. As shown in \autoref{map}(a), the peak position exhibits a continuous and stochastic wander over several hours. To distinguish the underlying physical mechanism, we extracted the fluctuation rates using autocorrelation function (ACF) analysis across various laser excitation powers [\autoref{map}(b)]. When the optical power changes by one order of magnitude, the increase in the fluctuation rate remains less than threefold [\autoref{map}(c)], implying that these fluctuations are more likely driven by a thermally-activated background bath rather than single/two - photon process. More analysis and data are provided in Supplementary Material Section S2.5-S2.6 \cite{SM}. To evaluate the applicability of our method for material characterization, we investigated two 4H-SiC epitaxial wafers (Wafer 1 and Wafer 2) obtained from different commercial manufacturers. Both samples were processed using identical ion implantation and annealing protocols to generate the PL5 center probes (details in Supplementary Material Section S1.3 \cite{SM}). Despite sharing the same ion implantation process, the sensors exhibited strikingly different noise environments. While wafer 1 showed widespread spectral diffusion and instability [\autoref{map}(d)], wafer 2 demonstrated high spectral stability with minimal near DC electrical noise [\autoref{map}(f)]. This pronounced contrast demonstrates that our method is sufficiently sensitive to reveal the intrinsic differences in charge traps density and material quality between two wafers. By correlating the spatial locations of individual sensors with their measured noise levels, quantified by the standard deviation of the ODMR peak positions ($\sigma_f$), we constructed electrical noise maps for both samples [\autoref{map}(e,g)]. The noise map of Wafer 1 reveals macroscopic spatial variations in noise density over tens of micrometers and the distribution of fluctuation rate across different PL5 sensors extracted by ACF method is provided in Supplementary Material Section S2.5 \cite{SM} . This demonstrates the potential of our method as a powerful characterization tool for imaging nanoscale electric noise distributions with single-charge sensitivity and wafer-quality diagnostics. Furthermore, conventional electrical characterization and ensemble defect measurements on both wafers \cite{SM} yield results consistent with the disparities observed via our quantum sensors (Supplementary Material Section S4). This agreement supports the reliability of the PL5-based measurements, while the single-defect quantum probes provide complementary nanoscale insight into local charge-noise heterogeneity beyond the wafer-averaged response.
 \begin{figure*}[!htbp]
\includegraphics{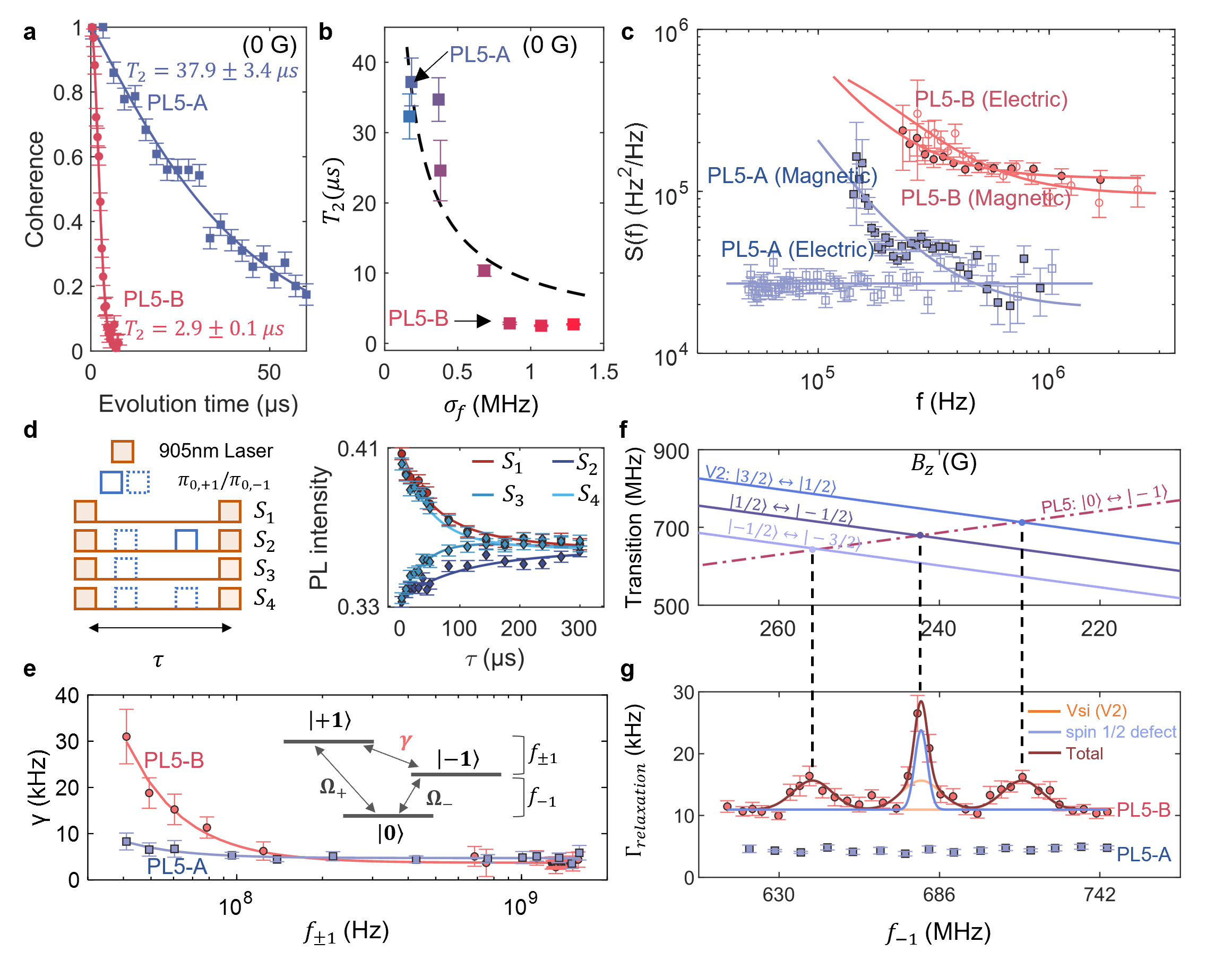}
\caption{\justifying \textbf{Noise strength correlation and spectrum of PL5 centers with EPR identification of noise source.} (a) Hahn echo $T_2$ measurement of the two PL5 centers (named PL5-A and PL5-B) on wafer 1. The sensor with lower near-DC noise (PL5-A) exhibits a longer coherence time. (b) A rapid decrease in Hahn echo coherence time is observed as the spectral fluctuation ($\sigma_f$) increases, indicating a strong correlation between near-DC noise and high-frequency electrical noise. The dashed cureve serves as a fitting guideline with $T_2\propto 1/\sigma^k$. (c) Noise spectra obtained from XY8 dynamical decoupling measurements on PL5-A and PL5-B. Experiments were performed under both zero field and 120G magnetic field aligned along the defect axis, fit with Lorentz model.
(d) SQ and DQ transition rates are extracted using different spin state initialization and readout schemes. Right pannel shows the measurement result of PL5-B at 260.4G along PL5 axis. Fitting model follows the equation shown in Eq.\ref{relaxation}. 
(e) Comparison of electrical noise strength $\gamma$ between the two PL5 centers, showing that the unstable sensor exhibits a stronger 1/$f$-type noise.
(f) Transition energy ($\Delta$) for PL5 center and V2 center are plotted against magnetic field. (g) $T_1$ relaxometry for two PL5 centers. Here, relaxation rate $\Gamma$ was extracted from $S_1-S_3$ shown in (d). The observed accelerated relaxation of PL5-B reveals the presence of environmental noise.  Based on the extracted spectral peak position and calculation of transition energy shown in (f), we attribute this signal to nearby silicon vacancy (V2) and other spin-1/2 traps, while no signal was observed for PL5-A. Errorbar is given by 1 SD.}
\label{noisespectra}
\end{figure*}

\textbf{Noise Spectrum and EPR Identification}
Having visualized the spatial heterogeneity of the electric noise, we carried out a detailed study on Sample 1 in subsequent experiments. We sought to identify its microscopic origin and explore the correlation in frequency domain. First, we measured the Hahn echo coherence time $T_2$ of 8 sensors. Among these, two representative PL5 centers (designated PL5-A and PL5-B) exhibiting different CW-ODMR stability ($\sigma_{fA}$ = 0.18 MHz; $\sigma_{fB}$ = 0.86 MHz) were selected for detailed investigation. Note that here we use the two-point measurement method to enhance the temporal resolution \cite{SM,2015-TimeResolvedLuminescenceNanothermometry-Tzeng-NanoLett.,2011-RealTimeMagnetic-Schoenfeld-Phys.Rev.Lett.}. As shown in \autoref{noisespectra}(a), the unstable sensor PL5-B exhibits a pronounced reduction in $T_2$ by up to an order of magnitude compared to the stable sensor PL5-A. To more clearly reveal this correlation, we plotted $\sigma_f$ of the 8 sensors against their Hahn echo coherence times as shown in \autoref{noisespectra}(b).The observed trend indicates that PL5 centers experiencing stronger near-DC spectral diffusion generally exhibit shorter Hahn-echo coherence times. The dashed line is fitted with $T_2 \propto 1/\sigma^k$ $(k \approx 0.8)$, an empirical power-law fit to the sensors under the assumption of a universal noise spectral density across the sample. However, for the three PL5 centers with the largest values of $\sigma_f$, the measured coherence times fall below the fitted trend. These experimental deviations suggest that the local defect environments around these PL5 centers may exhibit distinct dynamical behaviors. To further investigate the noise spectra of PL5 A and B at higher frequencies up to MHz, we implemented XY8-1 dynamical decoupling protocols on the two PL5 under zero-field and axial magnetic field ($\sim$120 G) with the resulting noise spectrum analyzed using Lorentzian fits \cite{2017-DoubleQuantumSpinRelaxationLimits-Myers-Phys.Rev.Lett.}. At zero magnetic field, longitudinal magnetic noise is suppressed by the transverse strain field $E_x$, so decoherence is governed by electric noise along x direction. In contrast, under $B_z >$  100 G, the transverse electromagnetic fluctuations are suppressed, and because the Stark coefficient for the longitudinal electric response of PL5 center is exceptionally small—longitudinal magnetic noise becomes the dominant source of decoherence \cite{2017-DoubleQuantumSpinRelaxationLimits-Myers-Phys.Rev.Lett.,PhysRevApplied.13.034010}. Under the delta-function filter approximation, we obtained the noise spectrum as shown in \autoref{noisespectra}(c). The results indicate a difference in the noise environments of the two sensors. For the spectrally stable PL5-A at near-DC range (with lower $\sigma_f$), the electric noise spectrum also stays at a low level in the sub-MHz range, implying that its magnetic noise is predominantly caused by the surrounding spin bath. In contrast, PL5-B exhibits significantly higher electric noise, accompanied by a magnetic noise intensity several times greater than that of PL5-A. This suggests that, beyond the intrinsic spin bath, PL5-B is coupled to local spin-active or paramagnetic fluctuators whose charge-state dynamics are correlated with the enhanced electric-noise environment.
These findings confirm that minimizing the proximity effect of electromagnetic noise through fundamental charge engineering is a critical step to realizing high-coherence quantum sensors \cite{zeledon2025minutelongquantumcoherenceenabled,2019-ElectricalOpticalControl-Anderson-Science}. More results of higher order dynamical decoupling are provided in the Supplementary Materials Section S3.1 \cite{SM}.

To extend our noise-spectroscopy window from the MHz to GHz regime, we applied a magnetic field along the PL5 center axis and scanned the $B_z$ to tune the PL5 centers energy levels and subsequently performed $T_1$ relaxation measurement. By measuring the evolution curves under different initialization and readout conditions $S_1\sim S_4$ as shown in [\autoref{noisespectra}(d)], we extracted the single-quantum (SQ) and double-quantum (DQ) transition rates, denoted as $\Omega_+/\Omega_-$ and $ \gamma$ as shown in \autoref{noisespectra}(e) (inset figure) \cite{2017-DoubleQuantumSpinRelaxationLimits-Myers-Phys.Rev.Lett.}, which represent the magnetic and electric noise strength in our system. After preparing the initial state of PL5 center at the eigenstates  $m_s = 0, \pm 1$ by laser and microwave pulses ($\pi_{0,+1},\pi_{0,-1}$), the process of longitudinal relaxation of spin state $\rho (t)$ can be described by rate equation
\begin{equation}\dot{\rho}(t)=\Gamma\rho(t)
\label{relaxation}\end{equation}
where the decay matrix is given by \cite{2017-DoubleQuantumSpinRelaxationLimits-Myers-Phys.Rev.Lett.,10.1093/nsr/nwad100}
\begin{equation}\Gamma=\begin{pmatrix}-\Omega_{+1}-\gamma&\Omega_{+1}&\gamma\\\Omega_{+1}&-\Omega_{+1}-\Omega_{-1}&\Omega_{-1}\\\gamma&\Omega_{-1}&-\Omega_{-1}-\gamma\end{pmatrix}
\end{equation}

\begin{figure*}[htbp]
\includegraphics[scale=0.88]{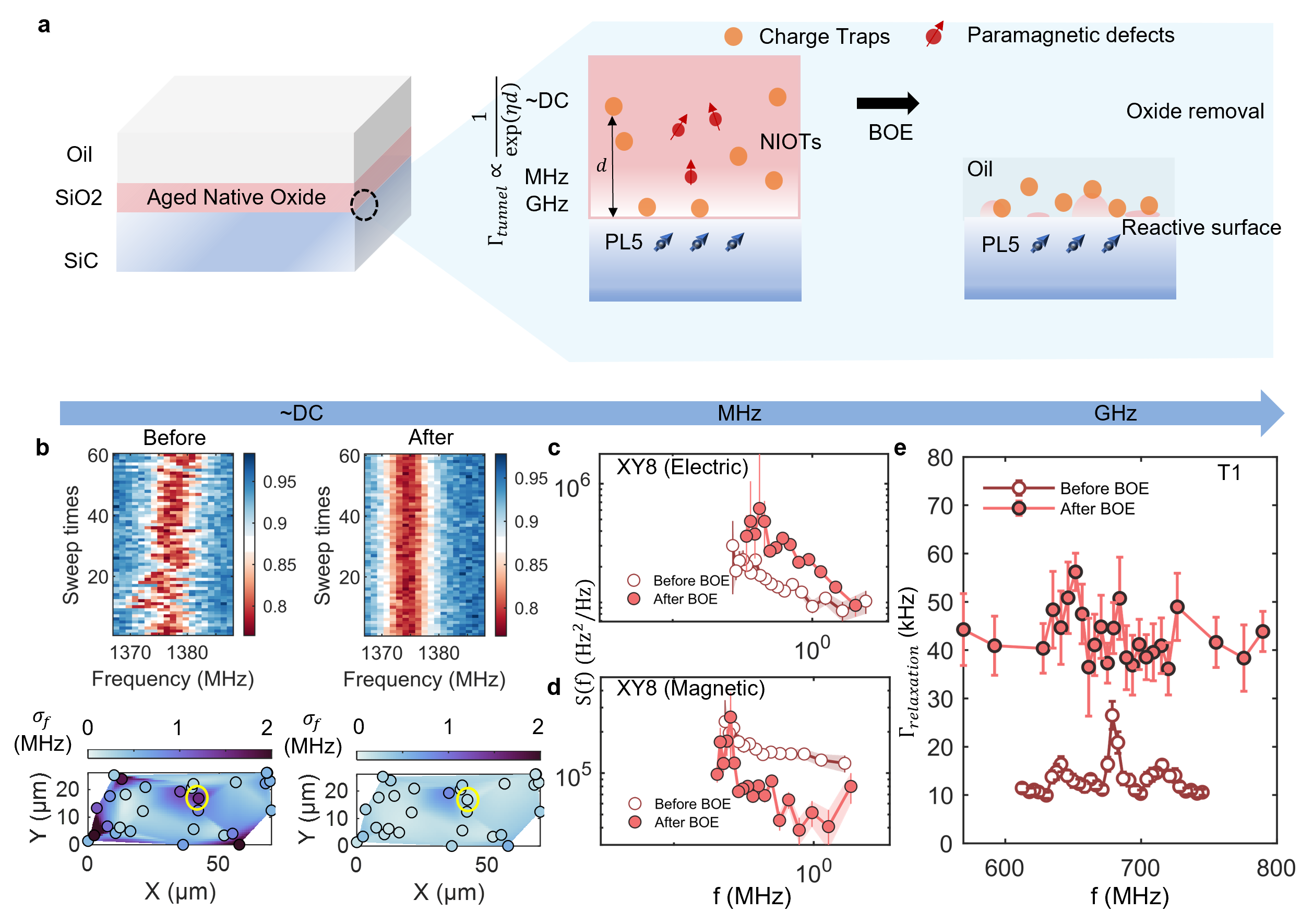}
\caption{\justifying \textbf{Validation of the interface spectral noise model via surface chemical engineering.} (a) Schematic of the SiC-SiO$2$ interface architecture. The charge tunneling rate $(\Gamma_{tunnel})$ is exponentially dependent on the trap depth $d$. Buffered oxide etch (BOE) strips the aged native oxide and its near DC noise contribution of near-interface oxide traps (NIOTs), but exposes a reactive bare SiC surface that rapidly accumulates unpassivated surface states (e.g.,dangling bond). (b) Near-DC noise stabilization. Continuous ODMR tracking (top) and corresponding spatial noise maps (bottom) show a global suppression of the slow spectral diffusion ($\sigma_f$) after BOE treatment in sample 1. (c-e) Comparison of broadband noise decoupling result and $T_1$ collapse for the unstable sensor PL5-B. In the MHz regime, the electric noise at 0 G (c) increases due to the highly active bare surface (more charge traps), whereas the magnetic noise at 120 G (d) decreases due to the removal of the oxide paramagnetic spin bath. In the GHz regime (e), the $T_1$ relaxation rate ($\Gamma_{relaxation}$) increase significantly, obscuring the previously resolved EPR resonances.}
\label{boe}
\end{figure*}

Based on this method, we observed two important phenomena. First, in the tens-of-MHz regime of DQ transition rate measurements, we observed $1/f^\alpha$-type electrical noise in both PL5-A and PL5-B as shown in \autoref{noisespectra}(e). The PL5-B, which shows unstable ODMR positions in former experiment, exhibits stronger $1/f^\alpha$ noise compared to PL5-A. This confirms the presence of an ensemble of charge fluctuators in the environment surrounding the unstable quantum sensor, which provide valuable insight into the characterization of the noise environment in semiconductors. Secondly, we examined its magnetic noise spectrum by varying the magnetic field and extracted relaxation rate $\Gamma$ from $S_1-S_3$ shown in \autoref{noisespectra}(d). The resulting spectra reveal a direct causal link between specific defect species and charge noise. For PL5-B, we observed distinct resonance peaks at $g=2.0$, accompanied by satellite features [\autoref{noisespectra}(f-g)], which can be attributed to relaxation enhancement between PL5-B and nearby traps. Data were fitted using a composite model comprising three Gaussian functions of equal amplitude and width, supplemented by an additional central Gaussian peak. Based on the fitting result ($f_1=641.58\pm 2.24, f_2=679.66\pm 0.28, f_3=714.94\pm 2.24$ MHz) and theoretical calculation [\autoref{noisespectra}(f)], our results corresponds to a defect configuration consisting of silicon vacancies ($\mathrm{V_{Si}}$(V2), $2D$ = 70 MHz) \cite{2017-IdentificationSivacancyRelated-Ivady-Phys.Rev.B} while the distinct relaxation peak emerging in the middle provides a signature of other other spin-1/2 dark traps background (e.g. carbon vacancies $\mathrm{V_{C}}$) as shown in \autoref{noisespectra}(g). In comparison, we observed no signal for the CW-stable quantum sensor PL5-A. When the $B_z$ was scanned over a wide range from 20 - 300 G, the SQ and DQ transition rates remained unchanged, maintaining a ratio $\frac{\gamma}{\Omega}\approx$ 2.4, where $\Omega \equiv \Omega_{+1}=\Omega_{-1}$ (see Supplementary Material Section S3.1 \cite{SM}). This result demonstrates the capability of our method to perform spatially resolved chemical identification. By retrieving spectroscopic fingerprints from distinct locations, we can effectively resolve variations in the local spin environment within different nanoscopic regions of the sample using a single quantum sensor. Finally, our estimation of the magnetic sensing range ($5\sim 8$ $\mathrm{nm}$) shows a spatial consistency with the electrical noise measurements and more results on different PL5 are provided in Supplementary Material Section S2.7 \cite{SM}.

\textbf{Native-Oxide Noise Origin Revealed by Surface Engineering}
To establish a physical microscopic origin for the observed broadband noise, we consider the architecture of the SiC-SiO$_2$ interface. Extensive semiconductor device studies indicate that the oxide inherently hosts a sub-stoichiometric SiO$_x$ transition layer \cite{2019-NanoscaleDepthControl-Li-Nanoscale} populated with near-interface oxide traps (NIOTs). Crucially, the charge tunneling dynamics of these defects are exponentially dependent on their physical distance from the semiconductor surface, generating a massive hierarchy of fluctuation timescales from nanoseconds to hours \cite{2019-CharacterizationSiO24HSiC-Fiorenza-Energies}. Such a temporal hierarchy can span from slow near-DC drift to nanosecond-scale fluctuations, making local broadband noise spectroscopy particularly important for understanding heterogeneous SiC/SiO2 interface quality and its impact on device-relevant noise. Here, we performed surface engineering using a buffered oxide etch (BOE, 7:1) to chemically strip the aged native oxide layer (\Cref{boe}(a)). This intervention fundamentally reconfigures the local defect landscape through a dual effect: it physically removes the thick outer oxide, while simultaneously exposing a highly reactive SiC surface that rapidly develops unpassivated, dynamic surface states and oxide islands upon brief exposure to the ambient environment. This reconfiguration results in a frequency-dependent decoupling of the quantum noise detected by the PL5 sensors. By stripping the thick native oxide, we effectively suppress the slow and intermediate noise channels.In the near-DC regime, spatial ODMR-tracking maps show a systematic reduction of slow spectral diffusion ($\sigma_f$) after BOE treatment, while a representative PL5 trace in \Cref{boe}(b) illustrates this stabilization at the single-sensor level. To examine how this surface modification affects different frequency channels in the same local environment, we further performed a detailed before/after comparison on the previously studied unstable sensor PL5-B. For PL5-B, $\sigma_f$ decreases from 0.86 MHz to 0.20 MHz. Similarly, in the intermediate MHz regime, the magnetic noise extracted at 120 G is significantly suppressed (\Cref{boe}(d))with an increase in the XY8 coherence time from $7.16 \pm 0.63~\mu$s to $12.57 \pm 0.61~\mu$s. This 1.76-fold enhancement indicates the quenching of the paramagnetic spin bath associated with the aged oxide. However, the newly exposed surface also introduces strong charge fluctuations, which elevate the high-frequency electric-noise channel and are accompanied by a pronounced GHz magnetic-noise background, likely arising from newly formed paramagnetic fluctuators near the reactive surface. Consequently, the overall $T_1$ relaxation rate ($\Gamma_{relaxation}$) exhibits a substantially increased background (\Cref{boe}(e)), obscuring the previously resolved EPR resonances signal. Ultimately, these divergent spectral trends illustrate how the structural depth of interfacial defects dictates the footprint of quantum decoherence. Details of the surface-treatment procedure and additional before/after comparisons for other PL5 centers are provided in Supplementary Material Section S2.7 and S3.2 \cite{SM}.

\section{Discussion}

In summary, our study establishes a robust, broadband quantum sensing protocol capable of diagnosing wafer-level quality variations with nanoscale precision by employing individual PL5 centers as room-temperature noise spectrometers. Through systematic monitoring, we demonstrate real-time observation of single-charge tunneling dynamics at room temperature in silicon carbide. Expanding from this microscopic insight,  we introduce a spatial imaging approach for noise characterization that provides a quantitative local indicator of wafer quality. The difference between wafers resolved by PL5 probes is further supported by conventional macroscopic characterization techniques, including CW-EPR, I-V/C-V, mercury-probe C-V, and DLTS, as detailed in Supplementary Material Section S4 \cite{SM}. These measurements provide wafer-level benchmarks that are consistent with the PL5 observations and help place the local quantum-sensing results in a broader materials-characterization context. Beyond mapping the noise, our broadband spectroscopy provides a frequency-resolved view of the local noise channels contributing to decoherence. By integrating dynamical decoupling and $T_1$ relaxometry, we provide the nanoscale EPR spectroscopic fingerprint of charge defects in SiC. The identification of distinct local spin species, such as silicon vacancies (V2) and other spin-1/2 defects, demonstrates dipole-dipole interactions among heterogeneous qubits (PL5 and V2). Given their distinct excitation and emission bands \cite{Castelletto_2024,2025-QuantumSensingSpin-Roberts-ACSNano}, this highlights the promising prospect of realizing scalable hybrid quantum registers within the same material platform. Furthermore, by combining broadband sensing with targeted surface chemical treatment (BOE), we identify a major contribution of the native oxide layer to the measured broadband noise. The before/after comparison supports a depth-dependent spectral picture: removing the oxide suppresses slow near-DC spectral diffusion and intermediate MHz magnetic noise from structurally constrained traps, but simultaneously exposes a highly reactive bare SiC surface that drives ultrafast charge fluctuations and a massive GHz magnetic noise background. Overall, individual quantum probes provide nanoscale access to local interfacial noise, enabling single-charge-level dynamics and frequency-resolved electric/magnetic-noise spectroscopy from near-DC to GHz regimes. This local, broadband perspective complements wafer-scale characterization by linking macroscopic material-quality variations to microscopic noise dynamics.

From the perspective of qubit optimization, our results highlight the role of substrate-hosted and interfacial charge environment in driving solid-state qubit decoherence. By identifying the frequency domains in which these fluctuators contribute, this work emphasizes the importance of high-quality substrates and controlled surface passivation for realizing high-performance and scalable quantum systems.

From the perspective of advanced metrology, our work demonstrates the capability of shallow PL5 centers as quantum probes for local interface diagnostics. Combined with near-infrared excitation, which minimizes above-band-gap photocarrier generation in 4H-SiC, this approach complements diamond NV-based sensing methods, especially in situations where external electric-field sensing can be affected by surface charge screening. Looking forward, commercial SiC wafers may be developed into smart quantum substrates or scanning probes for local multimodal diagnostics. Such a platform could be useful for monitoring diverse nanoscale environments, ranging from ferroelectric vortices in epitaxially grown van der Waals heterostructures \cite{2023-TwodimensionalVanWaals-Wang-Nat.Mater.} to interface Coulomb scattering in SiC-CMOS devices \cite{10.1063/5.0002838}.

This approach can be generalized to other defect species in SiC and wide-field imaging can be incorporated in practical measurement to increase speed and field of view \cite{monge2025ensembleaveragingparallelizedsingleshot}. This method is expected to serve as a vital tool for the precise characterization of third-generation semiconductors, paving the way for high-throughput, atomic scale defect engineering and spin-spin interfaces of qubits
implemented in silicon carbide for quantum information and sensing applications.

\section{Materials and Methods}

\textbf{Optical Setup.} A home-built scanning confocal microscope equipped with a three-axis piezoelectric stage (CoreMorrow Ltd, P15.XYZ300S-B) and a high numerical aperture (NA = 1.45) infrared oil-immersion objective (Nikon, CFI Plan Apochromat Lambda D) were employed in the experiments. Excitation of the PL5 was achieved using a 905-nm laser source (Changchun New Industries, MDL-III-905-SM). The fluorescence emission was further filtered through a longpass filter and then coupled into a single-mode fiber, guided to a superconducting nanowire single-photon detector (SNSPD) from Photon Technology (P-SPD-8S). Photon counts were recorded using a data acquisition counter (JYTEK, PCIe-5510).

\textbf{Coherent Manipulation Setup.} Microwave pulses were generated by a microwave signal generator (Rigol, DSG-3065B), followed by gating via a high-speed microwave switch (Mini-Circuits, ZASWA-2-50DR+). The signals were subsequently amplified using a broadband microwave amplifier (Mini-Circuits, ZHL-25W-272+) and delivered to a gold coplanar waveguide fabricated beneath the surface of the 4H-SiC sample. Synchronization of the optical pulses, microwave signals, and photon counting was precisely controlled by an digital delaypulse generator from CIQTEK.

\textbf{Sample Information.} In this study, quantum sensors were investigated in two samples, designated Sample 1 and Sample 2 in the main text. The two samples are from two different wafers purchased from different companies. Sample 1 was prepared by cutting from a wafer comprising a 12.5 $\mu$ m thick low-doped epitaxial layer of single-crystal 4H-SiC grown on a $4^\circ$ off-axis N-type 4H-SiC substrate. Similarly, Sample 2 was obtained from a wafer featuring a 30-$\mu$m-thick low-doped epitaxial layer of single-crystal 4H-SiC grown on a $4^\circ$ off-axis N-type 4H-SiC substrate. Both samples underwent ion implantation with 60-keV $^{14}\text{N}^{+}$ ions at a fluence of approximately $1\times 10^{10} \ \text{cm}^{-2}$. Post-implantation annealing was performed in a vacuum furnace at 1000 $^\circ$C for Sample 1 and 1050 $^\circ$C for Sample 2, each for a duration of 30 minutes. Crucially, while typical intrinsic defects (e.g., divacancies, PL1-4) generated by the 60 keV implantation reside deeper in the bulk, the PL5 color centers investigated here are structurally oxygen vacancy \cite{chen2026oxygenvacancyquantumspindefects,2026-HighYieldEngineeringIdentification-Hu-AdvancedMaterials}. To evaluate their depth profile, we performed Stopping and Range of Ions in Matter (SRIM) simulations \cite{2010-SRIMStoppingRange-Ziegler-NuclearInstrumentsandMethodsinPhysicsResearchSectionB:BeamInteractionswithMaterialsandAtoms,SM}, modeling the process with an assumed 1-nm-thick native oxide ($\text{SiO}_2$) layer on the SiC surface. As illustrated in Fig. S1, the primary intrinsic vacancies (Si and C vacancies) exhibit a broad distribution peaking approximately 80--100 nm below the surface. In contrast, the oxygen atoms recoiled from the native oxide layer are distributed significantly shallower, predominantly confined to the top few nanometers. Although the subsequent high-temperature annealing process may induce some spatial redistribution and defect migration, this initial recoil profile strongly suggests that the oxygen-vacancy-related PL5 complexes are fundamentally shallow quantum sensors in our 60-keV $^{14}\text{N}^{+}$ ions implantation, making them exquisitely sensitive to the silicon carbide interfacial environment.

\section{References}
\nocite{2007-DetectingSubstepsRotary-Zarrabi-,2016-GuidelinesReliableAnalysis-Puglisi-IEEETrans.Instrum.Meas.a,1982-LeastSquaresQuantization-Lloyd-IEEETrans.Inform.Theory,2024-FastCharacterizationOptically-Stone-J.Phys.Chem.C,1990-NewCompensatedCarrPurcell-Gullion-JournalofMagneticResonance1969,2012-ComparisonDynamicalDecoupling-Wang-Phys.Rev.B,2019-NonvanishingEffectDetuning-Lang-Phys.Rev.A,2000-CharacterizationParamagneticDefect-Macfarlane-J.Appl.Phys.,SiC-book,2002-ContinuouswavePulsedEPR-Mizuochi-Phys.Rev.B,2001-CarbonVacancyrelatedDefect-Son-Phys.Rev.B,2012-NegativeSystemCarbon-Son-Phys.Rev.Lett.,doi:https://doi.org/10.1002/9781118313534.ch5,1997-DeepLevelDefects-Hemmingsson-J.Appl.Phys.}
\bibliography{main,supplement-only}

@article{1982-LeastSquaresQuantization-Lloyd-IEEETrans.Inform.Theory,
  title = {Least Squares Quantization in {{PCM}}},
  author = {Lloyd, S.},
  year = {1982},
  month = mar,
  journal = {IEEE Transactions on Information Theory},
  volume = {28},
  number = {2},
  pages = {129--137},
  issn = {0018-9448},
  doi = {10.1109/TIT.1982.1056489},
  urldate = {2025-07-07},
  copyright = {https://ieeexplore.ieee.org/Xplorehelp/downloads/license-information/IEEE.html},
  langid = {english}
}

@inproceedings{2007-DetectingSubstepsRotary-Zarrabi-,
  title = {Detecting Substeps in the Rotary Motors of {{F}} o {{F}} 1 -{{ATP}} Synthase by {{Hidden Markov Models}}},
  booktitle = {Biomedical {{Optics}} ({{BiOS}}) 2007},
  author = {Zarrabi, N. and D{\"u}ser, M. G. and Reuter, R. and Dunn, S. D. and Wrachtrup, J. and B{\"o}rsch, M.},
  editor = {Enderlein, Jorg and Gryczynski, Zygmunt K.},
  year = {2007},
  month = feb,
  pages = {64440E},
  address = {San Jose, CA},
  doi = {10.1117/12.701001},
  urldate = {2024-03-04},
  langid = {english}
}

@article{2017-DoubleQuantumSpinRelaxationLimits-Myers-Phys.Rev.Lett.,
  title = {Double-{{Quantum Spin-Relaxation Limits}} to {{Coherence}} of {{Near-Surface Nitrogen-Vacancy Centers}}},
  author = {Myers, B. A. and Ariyaratne, A. and Jayich, A. C. Bleszynski},
  year = {2017},
  month = may,
  journal = {Physical Review Letters},
  volume = {118},
  number = {19},
  pages = {197201},
  issn = {0031-9007, 1079-7114},
  doi = {10.1103/PhysRevLett.118.197201},
  urldate = {2024-04-12},
  copyright = {http://link.aps.org/licenses/aps-default-license},
  langid = {english}
}

@article{2010-SRIMStoppingRange-Ziegler-NuclearInstrumentsandMethodsinPhysicsResearchSectionB:BeamInteractionswithMaterialsandAtoms,
  title = {{{SRIM}} -- {{The}} Stopping and Range of Ions in Matter (2010)},
  author = {Ziegler, James F. and Ziegler, M.D. and Biersack, J.P.},
  year = {2010},
  month = jun,
  journal = {Nuclear Instruments and Methods in Physics Research Section B: Beam Interactions with Materials and Atoms},
  volume = {268},
  number = {11-12},
  pages = {1818--1823},
  issn = {0168583X},
  doi = {10.1016/j.nimb.2010.02.091},
  urldate = {2025-07-29},
  copyright = {https://www.elsevier.com/tdm/userlicense/1.0/},
  langid = {english}
}

@article{2014-ElectricallyMechanicallyTunable-Falk-Phys.Rev.Lett.,
  title = {Electrically and {{Mechanically Tunable Electron Spins}} in {{Silicon Carbide Color Centers}}},
  author = {Falk, Abram L. and Klimov, Paul V. and Buckley, Bob B. and Iv{\'a}dy, Viktor and Abrikosov, Igor A. and Calusine, Greg and Koehl, William F. and Gali, {\'A}d{\'a}m and Awschalom, David D.},
  year = {2014},
  month = may,
  journal = {Physical Review Letters},
  volume = {112},
  number = {18},
  pages = {187601},
  issn = {0031-9007, 1079-7114},
  doi = {10.1103/PhysRevLett.112.187601},
  urldate = {2022-12-30},
  langid = {english}
}

@article{2024-FastCharacterizationOptically-Stone-J.Phys.Chem.C,
  title = {Fast {{Characterization}} of {{Optically Detected Magnetic Resonance Spectra}} via {{Data Clustering}}},
  author = {Stone, Dylan G. and Whitefield, Benjamin and Kianinia, Mehran and Bradac, Carlo},
  year = {2024},
  month = aug,
  journal = {The Journal of Physical Chemistry C},
  volume = {128},
  number = {31},
  pages = {13147--13154},
  issn = {1932-7447, 1932-7455},
  doi = {10.1021/acs.jpcc.4c03864},
  urldate = {2024-10-23},
  copyright = {https://doi.org/10.15223/policy-029},
  langid = {english}
}

@article{2011-ElectricfieldSensingUsing-Dolde-NaturePhys,
  title = {Electric-Field Sensing Using Single Diamond Spins},
  author = {Dolde, F. and Fedder, H. and Doherty, M. W. and N{\"o}bauer, T. and Rempp, F. and Balasubramanian, G. and Wolf, T. and Reinhard, F. and Hollenberg, L. C. L. and Jelezko, F. and Wrachtrup, J.},
  year = {2011},
  month = jun,
  journal = {Nature Physics},
  volume = {7},
  number = {6},
  pages = {459--463},
  issn = {1745-2473, 1745-2481},
  doi = {10.1038/nphys1969},
  urldate = {2023-05-03},
  langid = {english}
}

@article{2011-RealTimeMagnetic-Schoenfeld-Phys.Rev.Lett.,
  title = {Real {{Time Magnetic Field Sensing}} and {{Imaging Using}} a {{Single Spin}} in {{Diamond}}},
  author = {Schoenfeld, Rolf Simon and Harneit, Wolfgang},
  year = {2011},
  month = jan,
  journal = {Physical Review Letters},
  volume = {106},
  number = {3},
  pages = {030802},
  issn = {0031-9007, 1079-7114},
  doi = {10.1103/PhysRevLett.106.030802},
  urldate = {2025-04-09},
  copyright = {http://link.aps.org/licenses/aps-default-license},
  langid = {english}
}

@article{2011-RoomTemperatureCoherent-Koehl-Nature,
  title = {Room Temperature Coherent Control of Defect Spin Qubits in Silicon Carbide},
  author = {Koehl, William F. and Buckley, Bob B. and Heremans, F. Joseph and Calusine, Greg and Awschalom, David D.},
  year = {2011},
  month = nov,
  journal = {Nature},
  volume = {479},
  number = {7371},
  pages = {84--87},
  issn = {0028-0836, 1476-4687},
  doi = {10.1038/nature10562},
  urldate = {2022-09-26},
  langid = {english}
}

@article{2014-NanoscaleDetectionSingle-Dolde-Phys.Rev.Lett.,
  title = {Nanoscale {{Detection}} of a {{Single Fundamental Charge}} in {{Ambient Conditions Using}} the {{NV}} - {{Center}} in {{Diamond}}},
  author = {Dolde, Florian and Doherty, Marcus W. and Michl, Julia and Jakobi, Ingmar and Naydenov, Boris and Pezzagna, Sebastien and Meijer, Jan and Neumann, Philipp and Jelezko, Fedor and Manson, Neil B. and Wrachtrup, J{\"o}rg},
  year = {2014},
  month = mar,
  journal = {Physical Review Letters},
  volume = {112},
  number = {9},
  pages = {097603},
  issn = {0031-9007, 1079-7114},
  doi = {10.1103/PhysRevLett.112.097603},
  urldate = {2023-12-01},
  langid = {english}
}

@article{2016-OpticalPatterningTrapped-Jayakumar-NatCommun,
  title = {Optical Patterning of Trapped Charge in Nitrogen-Doped Diamond},
  author = {Jayakumar, Harishankar and Henshaw, Jacob and Dhomkar, Siddharth and Pagliero, Daniela and Laraoui, Abdelghani and Manson, Neil B. and Albu, Remus and Doherty, Marcus W. and Meriles, Carlos A.},
  year = {2016},
  month = aug,
  journal = {Nature Communications},
  volume = {7},
  number = {1},
  pages = {12660},
  issn = {2041-1723},
  doi = {10.1038/ncomms12660},
  urldate = {2023-04-09},
  langid = {english}
}

@article{2021-OpticalActivationDetection-Lozovoi-NatElectrona,
  title = {Optical Activation and Detection of Charge Transport between Individual Colour Centres in Diamond},
  author = {Lozovoi, Artur and Jayakumar, Harishankar and Daw, Damon and Vizkelethy, Gyorgy and Bielejec, Edward and Doherty, Marcus W. and Flick, Johannes and Meriles, Carlos A.},
  year = {2021},
  month = oct,
  journal = {Nature Electronics},
  volume = {4},
  number = {10},
  pages = {717--724},
  issn = {2520-1131},
  doi = {10.1038/s41928-021-00656-z},
  urldate = {2025-07-29},
  langid = {english}
}

@article{2024-CorrelatedSpectroscopyElectric-Delord-NanoLett.,
  title = {Correlated {{Spectroscopy}} of {{Electric Noise}} with {{Color Center Clusters}}},
  author = {Delord, Tom and Monge, Richard and Meriles, Carlos A.},
  year = {2024},
  month = jun,
  journal = {Nano Letters},
  volume = {24},
  number = {22},
  pages = {6474--6479},
  issn = {1530-6984, 1530-6992},
  doi = {10.1021/acs.nanolett.4c00222},
  urldate = {2025-07-29},
  copyright = {https://creativecommons.org/licenses/by/4.0/},
  langid = {english}
}

@misc{monge2025ensembleaveragingparallelizedsingleshot,
      title={Beyond ensemble averaging: Parallelized single-shot readout of hole capture in diamond}, 
      author={Richard Monge and Yuki Nakamura and Olaf Bach and Jason Shao and Artur Lozovoi and Alexander A. Wood and Kento Sasaki and Kensuke Kobayashi and Tom Delord and Carlos A. Meriles},
      year={2025},
      eprint={2507.11722},
      archivePrefix={arXiv},
      primaryClass={cond-mat.mes-hall},
      url={https://arxiv.org/abs/2507.11722}, 
}

@article{2015-TimeResolvedLuminescenceNanothermometry-Tzeng-NanoLett.,
  title = {Time-{{Resolved Luminescence Nanothermometry}} with {{Nitrogen-Vacancy Centers}} in {{Nanodiamonds}}},
  author = {Tzeng, Yan-Kai and Tsai, Pei-Chang and Liu, Hsiou-Yuan and Chen, Oliver Y. and Hsu, Hsiang and Yee, Fu-Goul and Chang, Ming-Shien and Chang, Huan-Cheng},
  year = {2015},
  month = jun,
  journal = {Nano Letters},
  volume = {15},
  number = {6},
  pages = {3945--3952},
  issn = {1530-6984, 1530-6992},
  doi = {10.1021/acs.nanolett.5b00836},
  urldate = {2025-04-09},
  langid = {english}
}

@article{2016-GuidelinesReliableAnalysis-Puglisi-IEEETrans.Instrum.Meas.a,
  title = {Guidelines for a {{Reliable Analysis}} of {{Random Telegraph Noise}} in {{Electronic Devices}}},
  author = {Puglisi, Francesco Maria and Pavan, Paolo},
  year = 2016,
  month = jun,
  journal = {IEEE Transactions on Instrumentation and Measurement},
  volume = {65},
  number = {6},
  pages = {1435--1442},
  issn = {0018-9456, 1557-9662},
  doi = {10.1109/TIM.2016.2518880},
  urldate = {2026-08-08},
  copyright = {https://ieeexplore.ieee.org/Xplorehelp/downloads/license-information/IEEE.html}
}

@article{2017-IdentificationSivacancyRelated-Ivady-Phys.Rev.B,
  title = {Identification of {{Si-vacancy}} Related Room-Temperature Qubits in 4 {{H}} Silicon Carbide},
  author = {Iv{\'a}dy, Viktor and Davidsson, Joel and Son, Nguyen Tien and Ohshima, Takeshi and Abrikosov, Igor A. and Gali, Adam},
  year = 2017,
  month = oct,
  journal = {Physical Review B},
  volume = {96},
  number = {16},
  pages = {161114},
  issn = {2469-9950, 2469-9969},
  doi = {10.1103/PhysRevB.96.161114},
  urldate = {2022-12-31},
  langid = {english}
}

@article{RevModPhys.89.035002,
  title = {Quantum sensing},
  author = {Degen, C. L. and Reinhard, F. and Cappellaro, P.},
  journal = {Rev. Mod. Phys.},
  volume = {89},
  issue = {3},
  pages = {035002},
  numpages = {39},
  year = {2017},
  month = {Jul},
  publisher = {American Physical Society},
  doi = {10.1103/RevModPhys.89.035002},
  url = {https://link.aps.org/doi/10.1103/RevModPhys.89.035002}
}

@article{2017-ReviewSiliconCarbide-She-IEEETrans.Ind.Electron.,
  title = {Review of {{Silicon Carbide Power Devices}} and {{Their Applications}}},
  author = {She, Xu and Huang, Alex Q. and Lucia, Oscar and Ozpineci, Burak},
  year = {2017},
  month = oct,
  journal = {IEEE Transactions on Industrial Electronics},
  volume = {64},
  number = {10},
  pages = {8193--8205},
  issn = {0278-0046, 1557-9948},
  doi = {10.1109/TIE.2017.2652401},
  urldate = {2025-04-07},
  copyright = {https://ieeexplore.ieee.org/Xplorehelp/downloads/license-information/IEEE.html},
  langid = {american}
}

@article{2019-ElectricalOpticalControl-Anderson-Science,
  title = {Electrical and Optical Control of Single Spins Integrated in Scalable Semiconductor Devices},
  author = {Anderson, Christopher P. and Bourassa, Alexandre and Miao, Kevin C. and Wolfowicz, Gary and Mintun, Peter J. and Crook, Alexander L. and Abe, Hiroshi and Ul Hassan, Jawad and Son, Nguyen T. and Ohshima, Takeshi and Awschalom, David D.},
  year = {2019},
  month = dec,
  journal = {Science},
  volume = {366},
  number = {6470},
  pages = {1225--1230},
  issn = {0036-8075, 1095-9203},
  doi = {10.1126/science.aax9406},
  urldate = {2023-03-15},
  langid = {english}
}

@article{2020-SolutionElectricField-Oberg-Phys.Rev.Applied,
  title = {Solution to {{Electric Field Screening}} in {{Diamond Quantum Electrometers}}},
  author = {Oberg, L.M. and De Vries, M.O. and Hanlon, L. and Strazdins, K. and Barson, M S.J. and Doherty, M.W. and Wrachtrup, J.},
  year = {2020},
  month = jul,
  journal = {Physical Review Applied},
  volume = {14},
  number = {1},
  pages = {014085},
  issn = {2331-7019},
  doi = {10.1103/PhysRevApplied.14.014085},
  urldate = {2024-01-19},
  langid = {english}
}

@article{2022-CharacterizationMethodsDefects-Bathen-J.Appl.Phys.b,
  title = {Characterization Methods for Defects and Devices in Silicon Carbide},
  author = {Bathen, M. E. and Lew, C. T.-K. and Woerle, J. and Dorfer, C. and Grossner, U. and Castelletto, S. and Johnson, B. C.},
  year = {2022},
  month = apr,
  journal = {Journal of Applied Physics},
  volume = {131},
  number = {14},
  pages = {140903},
  issn = {0021-8979, 1089-7550},
  doi = {10.1063/5.0077299},
  urldate = {2025-04-07},
  langid = {english}
}

@article{2022-DefectInspectionTechniques-Chen-NanoscaleResLett,
  title = {Defect {{Inspection Techniques}} in {{SiC}}},
  author = {Chen, Po-Chih and Miao, Wen-Chien and Ahmed, Tanveer and Pan, Yi-Yu and Lin, Chun-Liang and Chen, Shih-Chen and Kuo, Hao-Chung and Tsui, Bing-Yue and Lien, Der-Hsien},
  year = {2022},
  month = dec,
  journal = {Nanoscale Research Letters},
  volume = {17},
  number = {1},
  pages = {30},
  issn = {1556-276X},
  doi = {10.1186/s11671-022-03672-w},
  urldate = {2025-04-07},
  langid = {english}
}

@article{2022-ImagingDarkCharge-Lozovoi-Sci.Adv.,
  title = {Imaging Dark Charge Emitters in Diamond via Carrier-to-Photon Conversion},
  author = {Lozovoi, Artur and Vizkelethy, Gyorgy and Bielejec, Edward and Meriles, Carlos A.},
  year = {2022},
  month = jan,
  journal = {Science Advances},
  volume = {8},
  number = {1},
  pages = {eabl9402},
  issn = {2375-2548},
  doi = {10.1126/sciadv.abl9402},
  urldate = {2023-07-31},
  langid = {english}
}

@article{2022-RoomtemperatureCoherentManipulation-Li-Natl.Sci.Rev.,
  title = {Room-Temperature Coherent Manipulation of Single-Spin Qubits in Silicon Carbide with a High Readout Contrast},
  author = {Li, Qiang and Wang, Jun-Feng and Yan, Fei-Fei and Zhou, Ji-Yang and Wang, Han-Feng and Liu, He and Guo, Li-Ping and Zhou, Xiong and Gali, Adam and Liu, Zheng-Hao and Wang, Zu-Qing and Sun, Kai and Guo, Guo-Ping and Tang, Jian-Shun and Li, Hao and You, Li-Xing and Xu, Jin-Shi and Li, Chuan-Feng and Guo, Guang-Can},
  year = {2022},
  month = jun,
  journal = {National Science Review},
  volume = {9},
  number = {5},
  pages = {nwab122},
  issn = {2095-5138, 2053-714X},
  doi = {10.1093/nsr/nwab122},
  urldate = {2022-10-11},
  langid = {english}
}

@misc{chen2026oxygenvacancyquantumspindefects,
      title={Oxygen-vacancy quantum spin defects in silicon carbide}, 
      author={Yu Chen and Qi Zhang and Mingzhe Liu and Junda Wu and Jinpeng Liu and Xin Zhao and Jingyang Zhou and Pei Yu and Shaochun Lin and Yuanhong Teng and Wancheng Yu and Ya Wang and Changkui Duan and Fazhan Shi},
      year={2026},
      eprint={2504.07558},
      archivePrefix={arXiv},
      primaryClass={cond-mat.mtrl-sci},
      url={https://arxiv.org/abs/2504.07558}, 
}

@misc{SM,
    note =  {See Supplementary Material for details about CW-ODMR analysis, charge location calculation and other noise characteriztion results.}
}

@article{2024-3DMappingManipulationPhotocurrent-Wood-AdvancedMaterials,
  title = {{{3D}}-{{Mapping}} and {{Manipulation}} of {{Photocurrent}} in an {{Optoelectronic Diamond Device}}},
  author = {Wood, Alexander A. and McCloskey, Daniel J. and Dontschuk, Nikolai and Lozovoi, Artur and Goldblatt, Russell M. and Delord, Tom and Broadway, David A. and Tetienne, Jean-Philippe and Johnson, Brett C. and Mitchell, Kaih T. and Lew, Christopher T.-K. and Meriles, Carlos A. and Martin, Andy M.},
  year = {2024},
  month = aug,
  journal = {Advanced Materials},
  pages = {2405338},
  issn = {0935-9648, 1521-4095},
  doi = {10.1002/adma.202405338},
  urldate = {2024-08-26},
  langid = {english}
}

@article{2024-AtomicOpticalAntennas-Li-Nat.Photon.,
  title = {Atomic Optical Antennas in Solids},
  author = {Li, Zixi and Guo, Xinghan and Jin, Yu and Andreoli, Francesco and Bilgin, Anil and Awschalom, David D. and Delegan, Nazar and Heremans, F. Joseph and Chang, Darrick and Galli, Giulia and High, Alexander A.},
  year = {2024},
  month = jun,
  journal = {Nature Photonics},
  issn = {1749-4885, 1749-4893},
  doi = {10.1038/s41566-024-01456-5},
  urldate = {2024-07-10},
  langid = {english}
}

@article{2025-QuantumElectrometerTimeresolved-Pieplow-NatureCommunicationsa,
  title = {Quantum Electrometer for Time-Resolved Material Science at the Atomic Lattice Scale},
  author = {Pieplow, Gregor and Torun, Cem G{\"u}ney and Gurr, Charlotta and Munns, Joseph H. D. and Herrmann, Franziska Marie and Thies, Andreas and Pregnolato, Tommaso and Schr{\"o}der, Tim},
  year = {2025},
  month = jul,
  journal = {Nature Communications},
  volume = {16},
  number = {1},
  pages = {6435},
  issn = {2041-1723},
  doi = {10.1038/s41467-025-61839-2}
}

@article{2024-CorrelatedSensingSolidstate-Ji-Nat.Photon.,
  title = {Correlated Sensing with a Solid-State Quantum Multisensor System for Atomic-Scale Structural Analysis},
  author = {Ji, Wentao and Liu, Zhaoxin and Guo, Yuhang and Hu, Zhihao and Zhou, Jingyang and Dai, Siheng and Chen, Yu and Yu, Pei and Wang, Mengqi and Xia, Kangwei and Shi, Fazhan and Wang, Ya and Du, Jiangfeng},
  year = {2024},
  month = mar,
  journal = {Nature Photonics},
  volume = {18},
  number = {3},
  pages = {230--235},
  issn = {1749-4885, 1749-4893},
  doi = {10.1038/s41566-023-01352-4},
  urldate = {2024-11-03},
  langid = {english}
}

@article{2024-RoomtemperatureWaveguideIntegrated-Hu-NatCommun,
  title = {Room-Temperature Waveguide Integrated Quantum Register in a Semiconductor Photonic Platform},
  author = {Hu, Haibo and Zhou, Yu and Yi, Ailun and Bao, Tongyuan and Liu, Chengying and Luo, Qi and Zhang, Yao and Wang, Zi and Li, Qiang and Lu, Dawei and Liu, Zhengtong and Xiao, Shumin and Ou, Xin and Song, Qinghai},
  year = {2024},
  month = nov,
  journal = {Nature Communications},
  volume = {15},
  number = {1},
  pages = {10256},
  issn = {2041-1723},
  doi = {10.1038/s41467-024-54606-2},
  urldate = {2025-03-10},
  langid = {english}
}

@article{2024-SinglemoleculeScaleMagnetic-Du-Rev.Mod.Phys.,
  title = {Single-Molecule Scale Magnetic Resonance Spectroscopy Using Quantum Diamond Sensors},
  author = {Du, Jiangfeng and Shi, Fazhan and Kong, Xi and Jelezko, Fedor and Wrachtrup, J{\"o}rg},
  year = {2024},
  month = may,
  journal = {Reviews of Modern Physics},
  volume = {96},
  number = {2},
  pages = {025001},
  issn = {0034-6861, 1539-0756},
  doi = {10.1103/RevModPhys.96.025001},
  urldate = {2025-04-13},
  langid = {english}
}

@article{2024-TrackingSingleCharge-Weng-Phys.Rev.B,
  title = {Tracking Single Charge Jumping with an Individual Color Center in Diamond},
  author = {Weng, C. F. and Lin, S. R. and Zhao, J. X. and Yang, Y. J. and Cai, M. Y. and Guo, Y. H. and Lou, L. R. and Zhu, W. and Wang, G. Z.},
  year = {2024},
  month = jun,
  journal = {Physical Review B},
  volume = {109},
  number = {22},
  pages = {224104},
  issn = {2469-9950, 2469-9969},
  doi = {10.1103/PhysRevB.109.224104},
  urldate = {2024-08-06},
  langid = {english}
}

@article{PhysRevApplied.13.034010,
  title = {Fast Relaxation on Qutrit Transitions of Nitrogen-Vacancy Centers in Nanodiamonds},
  author = {Gardill, A. and Cambria, M.C. and Kolkowitz, S.},
  journal = {Phys. Rev. Appl.},
  volume = {13},
  issue = {3},
  pages = {034010},
  numpages = {8},
  year = {2020},
  month = {Mar},
  publisher = {American Physical Society},
  doi = {10.1103/PhysRevApplied.13.034010},
  url = {https://link.aps.org/doi/10.1103/PhysRevApplied.13.034010}
}

@article{10.1093/nsr/nwad100,
    author = {Zhao, Zhiyuan and Ye, Xiangyu and Xu, Shaoyi and Yu, Pei and Yang, Zhiping and Kong, Xi and Wang, Ya and Xie, Tianyu and Shi, Fazhan and Du, Jiangfeng},
    title = {Sub-nanotesla sensitivity at the nanoscale with a single spin},
    journal = {National Science Review},
    volume = {10},
    number = {12},
    pages = {nwad100},
    year = {2023},
    month = {04},
    issn = {2095-5138},
    doi = {10.1093/nsr/nwad100},
    url = {https://doi.org/10.1093/nsr/nwad100},
}

@article{2025-SingleV2Defect-Steidl-NatCommun,
  title = {Single {{V2}} Defect in {{4H}} Silicon Carbide {{Schottky}} Diode at Low Temperature},
  author = {Steidl, Timo and Kuna, Pierre and {Hesselmeier-H{\"u}ttmann}, Erik and Liu, Di and St{\"o}hr, Rainer and Knolle, Wolfgang and Ghezellou, Misagh and {Ul-Hassan}, Jawad and Schober, Maximilian and Bockstedte, Michel and Bian, Guodong and Gali, Adam and Vorobyov, Vadim and Wrachtrup, J{\"o}rg},
  year = {2025},
  month = may,
  journal = {Nature Communications},
  volume = {16},
  number = {1},
  publisher = {{Springer Science and Business Media LLC}},
  issn = {2041-1723},
  doi = {10.1038/s41467-025-59647-9},
  urldate = {2025-07-17},
  copyright = {https://creativecommons.org/licenses/by/4.0},
  langid = {english}
}

@misc{zeledon2025minutelongquantumcoherenceenabled,
      title={Minute-long quantum coherence enabled by electrical depletion of magnetic noise}, 
      author={Cyrus Zeledon and Benjamin Pingault and Jonathan C. Marcks and Mykyta Onizhuk and Yeghishe Tsaturyan and Yu-xin Wang and Benjamin S. Soloway and Hiroshi Abe and Misagh Ghezellou and Jawad Ul-Hassan and Takeshi Ohshima and Nguyen T. Son and F. Joseph Heremans and Giulia Galli and Christopher P. Anderson and David D. Awschalom},
      year={2025},
      eprint={2504.13164},
      archivePrefix={arXiv},
      primaryClass={quant-ph},
      url={https://arxiv.org/abs/2504.13164}, 
}

@article{2023-TwodimensionalVanWaals-Wang-Nat.Mater.,
  title = {Towards Two-Dimensional van Der {{Waals}} Ferroelectrics},
  author = {Wang, Chuanshou and You, Lu and Cobden, David and Wang, Junling},
  year = 2023,
  month = may,
  journal = {Nature Materials},
  volume = {22},
  number = {5},
  pages = {542--552},
  issn = {1476-1122, 1476-4660},
  doi = {10.1038/s41563-022-01422-y},
  urldate = {2025-12-04},
  langid = {english}
}

@article{10.1063/5.0002838,
  title = {An Advanced Physical Model for the {{Coulombic}} Scattering Mobility in {{4H-SiC}} Inversion Layers},
  author = {Naydenov, K. and Donato, N. and Udrea, F.},
  year = 2020,
  month = may,
  journal = {Journal of Applied Physics},
  volume = {127},
  number = {19},
  pages = {194504},
  issn = {0021-8979},
  doi = {10.1063/5.0002838}
}

@article{Castelletto_2024,
  title = {Quantum Systems in Silicon Carbide for Sensing Applications},
  author = {Castelletto, S and Lew, C T-K and Lin, Wu-Xi and Xu, Jin-Shi},
  year = 2023,
  month = dec,
  journal = {Reports on Progress in Physics},
  volume = {87},
  number = {1},
  pages = {014501},
  publisher = {IOP Publishing},
  doi = {10.1088/1361-6633/ad10b3}
}

@article{2025-QuantumSensingSpin-Roberts-ACSNano,
  title = {Quantum {{Sensing}} with {{Spin Defects Beyond Diamond}}},
  author = {Roberts, Henry and Abudayyeh, Hamza and Li, Xiaoqin and Li, Xiuling},
  year = 2025,
  month = jul,
  journal = {ACS Nano},
  volume = {19},
  number = {25},
  pages = {22528--22575},
  issn = {1936-0851, 1936-086X},
  doi = {10.1021/acsnano.5c00802},
  urldate = {2025-08-04},
  copyright = {https://doi.org/10.15223/policy-029},
  langid = {english}
}

@article{2019-CharacterizationSiO24HSiC-Fiorenza-Energies,
  title = {Characterization of {{SiO2}}/{{4H-SiC Interfaces}} in {{4H-SiC MOSFETs}}: {{A Review}}},
  shorttitle = {Characterization of {{SiO2}}/{{4H-SiC Interfaces}} in {{4H-SiC MOSFETs}}},
  author = {Fiorenza, Patrick and Giannazzo, Filippo and Roccaforte, Fabrizio},
  year = 2019,
  month = jun,
  journal = {Energies},
  volume = {12},
  number = {12},
  pages = {2310},
  issn = {1996-1073},
  doi = {10.3390/en12122310},
  urldate = {2026-04-26},
  langid = {english}
}

@article{2019-NanoscaleDepthControl-Li-Nanoscale,
  title = {Nanoscale Depth Control of Implanted Shallow Silicon Vacancies in Silicon Carbide},
  author = {Li, Qiang and Wang, Jun-Feng and Yan, Fei-Fei and Cheng, Ze-Di and Liu, Zheng-Hao and Zhou, Kun and Guo, Li-Ping and Zhou, Xiong and Zhang, Wei-Ping and Wang, Xiu-Xia and Huang, Wei and Xu, Jin-Shi and Li, Chuan-Feng and Guo, Guang-Can},
  year = 2019,
  journal = {Nanoscale},
  volume = {11},
  number = {43},
  pages = {20554--20561},
  issn = {2040-3364, 2040-3372},
  doi = {10.1039/C9NR05938E},
  urldate = {2022-10-08},
  langid = {english}
}

@article{1990-NewCompensatedCarrPurcell-Gullion-JournalofMagneticResonance1969,
  title = {New, Compensated {Carr-Purcell} Sequences},
  author = {Gullion, Terry and Baker, David B. and Conradi, Mark S.},
  year = {1990},
  journal = {Journal of Magnetic Resonance (1969)},
  volume = {89},
  number = {3},
  pages = {479--484},
  doi = {10.1016/0022-2364(90)90331-3}
}

@article{2012-ComparisonDynamicalDecoupling-Wang-Phys.Rev.B,
  title = {Comparison of Dynamical Decoupling Protocols for a Nitrogen-Vacancy Center in Diamond},
  author = {Wang, Zhi-Hui and {de Lange}, G. and Rist{\`e}, D. and Hanson, R. and Dobrovitski, V. V.},
  year = {2012},
  journal = {Physical Review B},
  volume = {85},
  number = {15},
  pages = {155204},
  doi = {10.1103/PhysRevB.85.155204}
}

@article{2019-NonvanishingEffectDetuning-Lang-Phys.Rev.A,
  title = {Nonvanishing Effect of Detuning Errors in Dynamical-Decoupling-Based Quantum Sensing Experiments},
  author = {Lang, J. E. and Madhavan, T. and Tetienne, J.-P. and Broadway, D. A. and Hall, L. T. and Teraji, T. and Monteiro, T. S. and Stacey, A. and Hollenberg, L. C. L.},
  year = {2019},
  journal = {Physical Review A},
  volume = {99},
  number = {1},
  pages = {012110},
  doi = {10.1103/PhysRevA.99.012110}
}

@article{2000-CharacterizationParamagneticDefect-Macfarlane-J.Appl.Phys.,
  title = {Characterization of Paramagnetic Defect Centers in Three Polytypes of Dry Heat Treated, Oxidized {SiC}},
  author = {Macfarlane, P. J. and Zvanut, M. E.},
  year = {2000},
  journal = {Journal of Applied Physics},
  volume = {88},
  number = {7},
  pages = {4122--4127},
  doi = {10.1063/1.1311810}
}

@book{SiC-book,
  title = {{SiC Power Materials}: Devices and Applications},
  editor = {Feng, Zhe Chuan and Hull, Robert and Osgood, R. M. and Parisi, J{\"u}rgen and Warlimont, Hans},
  year = {2004},
  series = {Springer Series in Materials Science},
  volume = {73},
  publisher = {Springer Berlin Heidelberg},
  address = {Berlin, Heidelberg},
  doi = {10.1007/978-3-662-09877-6}
}

@article{2002-ContinuouswavePulsedEPR-Mizuochi-Phys.Rev.B,
  title = {Continuous-Wave and Pulsed {EPR} Study of the Negatively Charged Silicon Vacancy in n-Type {4H-SiC}},
  author = {Mizuochi, N. and Yamasaki, S. and Takizawa, H. and Morishita, N. and Ohshima, T. and Itoh, H. and Isoya, J.},
  year = {2002},
  journal = {Physical Review B},
  volume = {66},
  number = {23},
  pages = {235202},
  doi = {10.1103/PhysRevB.66.235202}
}

@article{2001-CarbonVacancyrelatedDefect-Son-Phys.Rev.B,
  title = {Carbon Vacancy-Related Defect in {4H} and {6H SiC}},
  author = {Son, N. T. and Hai, P. N. and Janz{\'e}n, E.},
  year = {2001},
  journal = {Physical Review B},
  volume = {63},
  number = {20},
  pages = {201201},
  doi = {10.1103/PhysRevB.63.201201}
}

@article{2012-NegativeSystemCarbon-Son-Phys.Rev.Lett.,
  title = {Negative-{U} System of Carbon Vacancy in {4H-SiC}},
  author = {Son, N. T. and Trinh, X. T. and L{\o}vlie, L. S. and Svensson, B. G. and Kawahara, K. and Suda, J. and Kimoto, T. and Umeda, T. and Isoya, J. and Makino, T. and Ohshima, T. and Janz{\'e}n, E.},
  year = {2012},
  journal = {Physical Review Letters},
  volume = {109},
  number = {18},
  pages = {187603},
  doi = {10.1103/PhysRevLett.109.187603}
}

@inbook{doi:https://doi.org/10.1002/9781118313534.ch5,
  author = {Kimoto, Tsunenobu and Cooper, James A.},
  title = {Characterization Techniques and Defects in Silicon Carbide},
  booktitle = {Fundamentals of Silicon Carbide Technology},
  chapter = {5},
  pages = {125--187},
  publisher = {John Wiley \& Sons, Ltd},
  year = {2014},
  doi = {10.1002/9781118313534.ch5}
}

@article{1997-DeepLevelDefects-Hemmingsson-J.Appl.Phys.,
  title = {Deep Level Defects in Electron-Irradiated {4H SiC} Epitaxial Layers},
  author = {Hemmingsson, C. and Son, N. T. and Kordina, O. and Bergman, J. P. and Janz{\'e}n, E. and Lindstr{\"o}m, J. L. and Savage, S. and Nordell, N.},
  year = {1997},
  journal = {Journal of Applied Physics},
  volume = {81},
  number = {9},
  pages = {6155--6159},
  doi = {10.1063/1.364397}
}

@article{2026-HighYieldEngineeringIdentification-Hu-AdvancedMaterials,
  title = {High-Yield Engineering and Identification of Oxygen-Related Modified Divacancies in {4H-SiC}},
  author = {Hu, Qi-Cheng and Zhou, Ji-Yang and Ren, Shuo and He, Zhen-Xuan and Hao, Zhi-He and Liang, Rui-Jian and Lin, Wu-Xi and Han, Xiangru and Gali, Adam and Xu, Jin-Shi and Li, Chuan-Feng and Guo, Guang-Can},
  year = {2026},
  journal = {Advanced Materials},
  pages = {e73419},
  doi = {10.1002/adma.73419}
}

\textbf{Acknowledgments:} The authors thank Xun Li at the Shanghai Institute of Technical Physics (CAS) for assistance with electrical characterization measurements.

\textbf{Funding:} This work was supported by the National Natural Science Foundation of China (Grant Nos. T2125011), the CAS (Grant Nos. YSBR-068), Innovation Program for Quantum Science and Technology (Grant Nos. 2021ZD0302200, 2021ZD0303204, 2023ZD0300100), New Cornerstone Science Foundation through the XPLORER PRIZE, the Major Science and Technology Projects of Zhejiang Province (2025C01041)  and the Fundamental Research Funds for the Central Universities (226-2024-00142). This work was partially carried out at the USTC Center for Micro and Nanoscale Research and Fabrication, Nanofabrication facility in Suzhou Institute of Nanotech and Nanobionics (CAS) and the Shanghai Institute of Technical Physics (CAS).

\textbf{Author Contributions:} F.S. and Q.Z. supervised the project. Q.Z. conceived the idea. F.S., Q.Z. and J.L. designed the experiment. J.L., Y.C. and J.Y. prepared and built the setup. J.L., Y.T., Y.C., Y.-X.W., C.L., Q.L. and C.Y. performed the experiments and analyzed the data. J.L., Y.C., S.L. and Y.W. prepared the sample. J.L., Y.T., Q.Z. and F.S. wrote the manuscript. All authors discussed the results and commented on the manuscript.

\textbf{Competing Interests:} The authors declare that they have no competing interests.

\textbf{Data, Code, and Materials Availability:} All data and code needed to evaluate and reproduce the results in the paper are present in the paper and/or the Supplementary Materials. This study did not generate new materials.

\textbf{Supplementary Materials}

\noindent Supplementary Text Section S1 to S4\\
Fig S1 to S17\\
Table S1
\end{document}


\pagestyle{empty}
\makeatletter
\bibcite{2014-ElectricallyMechanicallyTunable-Falk-Phys.Rev.Lett.}{{24}{2014}{{Falk\ \emph  {et~al.}}}{{Falk, Klimov, Buckley, Iv{\'a}dy, Abrikosov, Calusine, Koehl, Gali,\ and\ Awschalom}}}
\bibcite{2007-DetectingSubstepsRotary-Zarrabi-}{{41}{2007}{{Zarrabi\ \emph  {et~al.}}}{{Zarrabi, D{\"u}ser, Reuter, Dunn, Wrachtrup,\ and\ B{\"o}rsch}}}
\bibcite{2016-GuidelinesReliableAnalysis-Puglisi-IEEETrans.Instrum.Meas.a}{{42}{2016}{{Puglisi}}{{}}}
\bibcite{1982-LeastSquaresQuantization-Lloyd-IEEETrans.Inform.Theory}{{43}{1982}{{Lloyd}}{{}}}
\bibcite{2024-FastCharacterizationOptically-Stone-J.Phys.Chem.C}{{44}{2024}{{Stone\ \emph  {et~al.}}}{{Stone, Whitefield, Kianinia,\ and\ Bradac}}}
\bibcite{1990-NewCompensatedCarrPurcell-Gullion-JournalofMagneticResonance1969}{{45}{1990}{{Gullion\ \emph  {et~al.}}}{{Gullion, Baker,\ and\ Conradi}}}
\bibcite{2012-ComparisonDynamicalDecoupling-Wang-Phys.Rev.B}{{46}{2012}{{Wang\ \emph  {et~al.}}}{{Wang, {de Lange}, Rist{\`e}, Hanson,\ and\ Dobrovitski}}}
\bibcite{2019-NonvanishingEffectDetuning-Lang-Phys.Rev.A}{{47}{2019}{{Lang\ \emph  {et~al.}}}{{Lang, Madhavan, Tetienne, Broadway, Hall, Teraji, Monteiro, Stacey,\ and\ Hollenberg}}}
\bibcite{2000-CharacterizationParamagneticDefect-Macfarlane-J.Appl.Phys.}{{48}{2000}{{Macfarlane\ and\ Zvanut}}{{}}}
\bibcite{SiC-book}{{49}{2004}{{Feng\ \emph  {et~al.}}}{{Feng, Hull, Osgood, Parisi,\ and\ Warlimont}}}
\bibcite{2002-ContinuouswavePulsedEPR-Mizuochi-Phys.Rev.B}{{50}{2002}{{Mizuochi\ \emph  {et~al.}}}{{Mizuochi, Yamasaki, Takizawa, Morishita, Ohshima, Itoh,\ and\ Isoya}}}
\bibcite{2001-CarbonVacancyrelatedDefect-Son-Phys.Rev.B}{{51}{2001}{{Son\ \emph  {et~al.}}}{{Son, Hai,\ and\ Janz{\'e}n}}}
\bibcite{2012-NegativeSystemCarbon-Son-Phys.Rev.Lett.}{{52}{2012}{{Son\ \emph  {et~al.}}}{{Son, Trinh, L{\o }vlie, Svensson, Kawahara, Suda, Kimoto, Umeda, Isoya, Makino, Ohshima,\ and\ Janz{\'e}n}}}
\bibcite{doi:https://doi.org/10.1002/9781118313534.ch5}{{53}{2014}{{Kimoto\ and\ Cooper}}{{}}}
\bibcite{1997-DeepLevelDefects-Hemmingsson-J.Appl.Phys.}{{54}{1997}{{Hemmingsson\ \emph  {et~al.}}}{{Hemmingsson, Son, Kordina, Bergman, Janz{\'e}n, Lindstr{\"o}m, Savage,\ and\ Nordell}}}

\makeatother

\title{Supplemental Material to ``Quantum Noise Spectroscopy of Nanoscale Charge Defects in Silicon Carbide at Room Temperature"}





\maketitle
\thispagestyle{empty}
\newpage

\tableofcontents 
\newpage

\section{Supplementary Sample Characterization and Data Overview}

 \begin{figure*}[htbp]
 \includegraphics{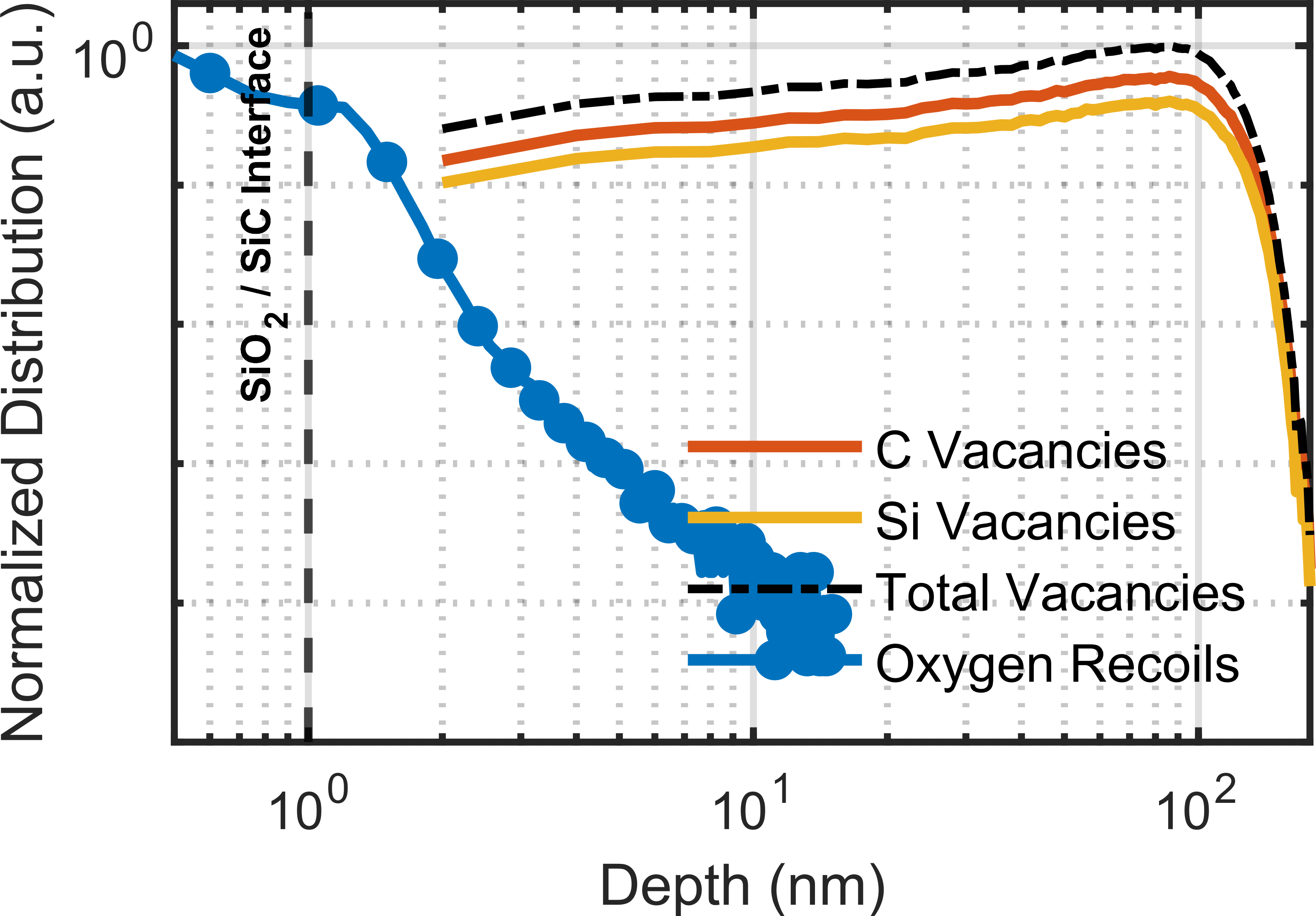}
 \caption{\label{figS: SRIM} Normalized depth profiles of oxygen recoils (blue circles) and lattice vacancies in 4H-SiC following a 60 keV $^{14}\text{N}^{+}$ ion implantation. The simulation assumes a 1-nm-thick native $\text{SiO}_2$ layer on the SiC surface. Unlike the intrinsic vacancies which peak deep in the epi layer (100nm), the oxygen recoils exhibit a drastically shallower spatial distribution. This difference confirms the near-surface nature of the oxygen-related PL5 color centers in our samples. } 
 \end{figure*}

\begin{figure*}[htbp]
 \includegraphics{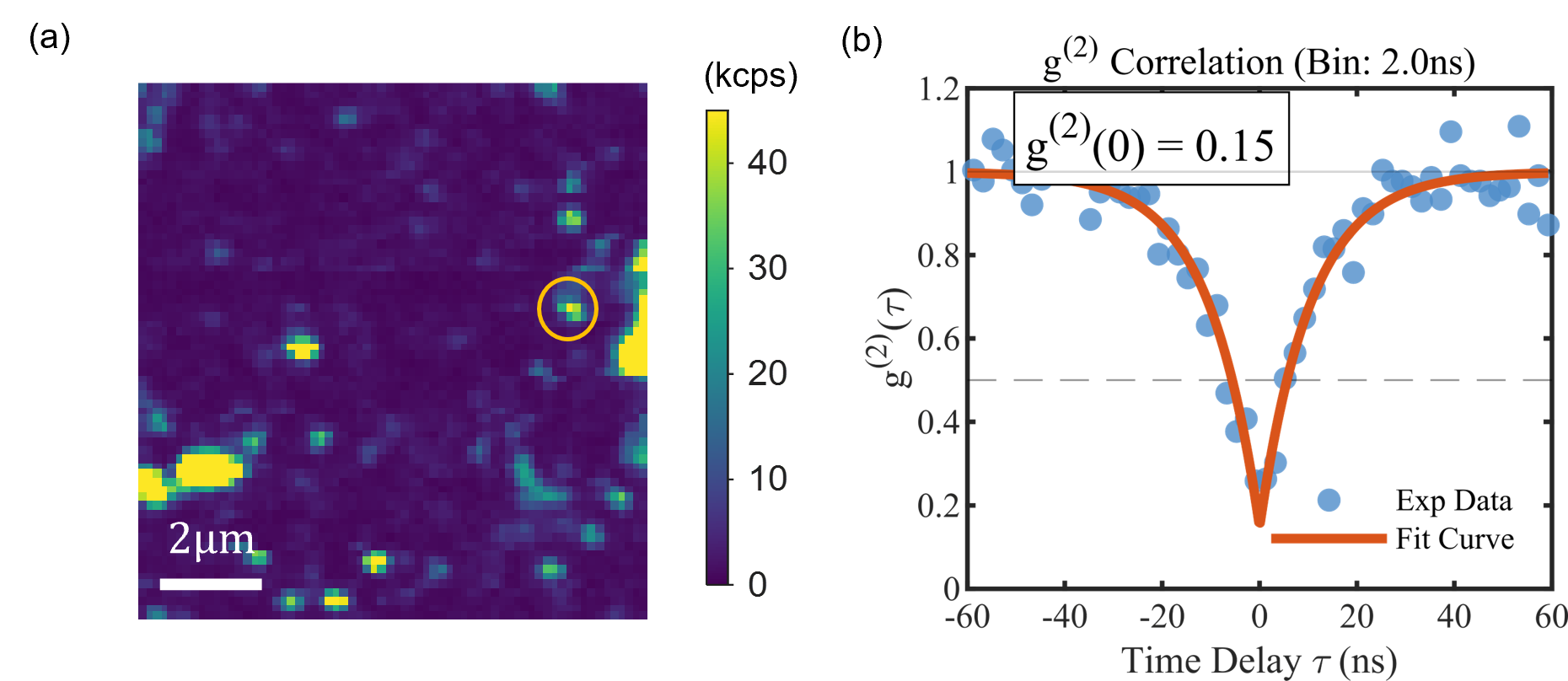}
 \caption{\label{figS: 单色心} (a) Confocal fluorescence image of the SiC sample, showing discrete fluorescent spots with good brightness uniformity. The orange circle indicates a representative color center selected for autocorrelation analysis. (b) Second-order autocorrelation $g^{(2)}(\tau)$ measurement of the circled spot in (a). The pronounced anti-bunching dip at zero delay, $g^{(2)}(0) = 0.15 < 0.5$, confirms that the emitter is a single color center  } 
 \end{figure*}

To provide a comprehensive overview of the quantum sensors investigated in this work and to clarify the statistical support for our claims, we summarize the PL5 centers used across different experimental figures in \Cref{tab:color_centers}. Besides, to verify the generation of single color centers using our current fabrication recipe, we characterized the sample using confocal photoluminescence microscopy. As shown in \Cref{figS: 单色心}(a), the confocal map reveals discrete fluorescent spots with a well-distributed brightness across the scanning area. To confirm the nature of these emitters, we performed a second-order photon autocorrelation measurement on a representative spot exhibiting typical PL5 brightness, as marked by the orange circle. The resulting curve  displays a pronounced anti-bunching dip with $g^{(2)}(0) = 0.15$, unambiguously demonstrating that the isolated spots generated under our current recipe are single color centers.
\begin{table*}[h]
    \centering
    \caption{\textbf{Summary of PL5 color centers investigated in this study.} This table outlines the specific color centers (identified by defect number, No.), the corresponding wafers, and the experiments performed to provide a clear mapping of the data presented in the main text figures.}
    \label{tab:color_centers}
    \begin{tabular}{c c c}
        \textbf{Experiment} & \textbf{Wafer (1/2)} & \textbf{Color center (PL5)}  \\
        \hline
        Single charge tracking (Fig.2) &  1 & No.1634 \\
        \hline
        Laser power dependency (Fig.3(a-c), Fig.S9) &  1 & No.328, 988 \\
        \hline
        DC noise statistics and mapping (Fig.3(d-g)) & 1 \& 2 & \makecell{64 color centers in Sample 1 \\ 63 color centers in Sample 2; \\ Representative traces: \\ No.24 in Sample 1, No.141 in Sample 2} \\
        \hline
        $T_2$ (Hahn echo) (Fig.4(b)) & 1 & \makecell{8 color centers (No.24, 120, 178 \\ 363, 385 (PL5-B), 423, 453 (PL5-A), 487)}  \\
        \hline
        \makecell{DD \& EPR relaxometry \\(Fig.4(a, c-g), Fig.5(c-e))} & 1 & \makecell{No.453 (PL5-A) \\ No.385 (PL5-B) } \\
        \hline
        \makecell{DC noise comparison \\ before/after BOE (Fig.5b)} & 1 & \makecell{30 color centers in Sample 1 \\ Representative traces: No.202}\\
        \hline
    \end{tabular}
\end{table*}

\newpage
\section{Analysis of ODMR Tracking} 
\label{secS: corr-map}
 \subsection{HMM Model and PCA Clustering}

 \begin{figure*}[htbp]
 \includegraphics{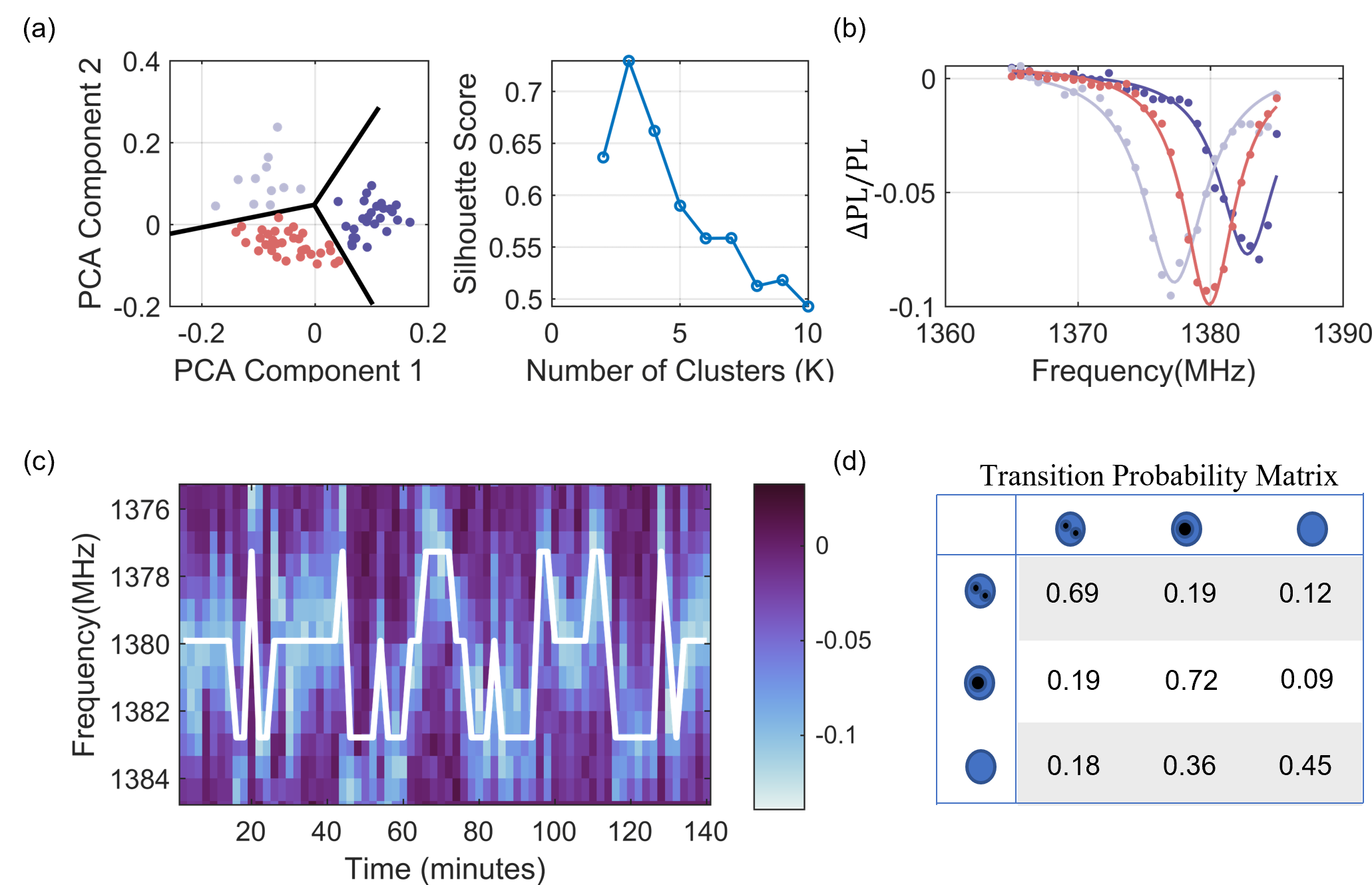}
 \caption{\label{figS: Cluster}(a) Clustering results visualized via principal component analysis (PCA), based on feature vectors constructed from contrast values at different frequencies. Silhouette coefficient is plotted as a function of cluster number, indicating optimal clustering performance at three clusters (K=3), where inter-class separability is maximized and intra-class variance is minimized. (b) Averaged CW spectral lineshapes corresponding to each of the three identified states. (c) Experimental CW spectral data and a three-state Markov process model based on the clustering result above. (d) Extracted transition constants for each state from the hidden Markov model, suggesting that the three discrete states may originate from surrounding dark defects with different charge configurations.  } 
 \end{figure*}
 
The behavior of dark defects capturing and releasing charges can be modeled using a hidden Markov model (HMM), with the capture and emission processes characterized by time constants $\tau_{release}$ and $\tau_{capture}$ respectively \cite{2007-DetectingSubstepsRotary-Zarrabi-,2016-GuidelinesReliableAnalysis-Puglisi-IEEETrans.Instrum.Meas.a}. In complex charge environments, it is often difficult to directly infer the number of hidden states from the spectral peak positions alone. For these color centers, fluctuations in the local electric field often occur within the timescale of a single CW-ODMR scan, leading to broadened or distorted spectral peaks, and in some cases, the appearance of multiple peaks. To better identify and classify these hidden spectral states, we treated each CW contrast curve as a feature vector and applied a clustering algorithm \cite{1982-LeastSquaresQuantization-Lloyd-IEEETrans.Inform.Theory,2024-FastCharacterizationOptically-Stone-J.Phys.Chem.C}. Principal component analysis (PCA) was used to reduce dimensionality and visualize the classification results, as shown in \autoref{figS: Cluster}(a). The optimal number of hidden states was determined using the silhouette coefficient, which indicates that three clusters provide the best separation for this dataset. The average contrast curves for each cluster are shown in \autoref{figS: Cluster}(b). We attribute the three observed states to different charge configurations of nearby dark charges. The transition rates between these states can be extracted using HMM analysis, as shown in \autoref{figS: Cluster}(c-d).

\subsection{CW Tracking between two color center}
\label{section:two color center}
\begin{figure*}[htbp]
\includegraphics{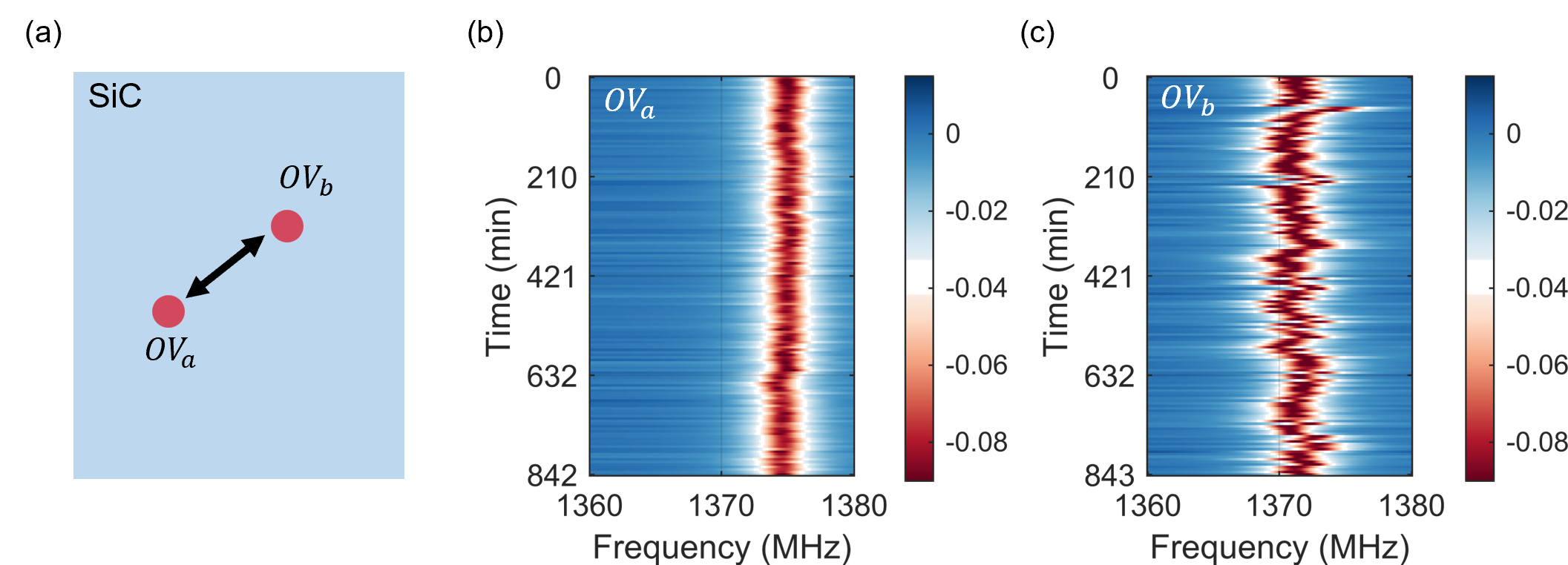}
\caption{\label{figS: 交叉}(a) Schematic of the experiment. We continuously alternate between two color centers, performing a CW-ODMR scan at the position of color center $OV_a$ to obtain the spectrum, then immediately move to color center $OV_b$ for measurement, before returning to $OV_a$. This process repeats in a cycle. (b-c) Results obtained by separating the experimental data from the two color centers. A clear distinction is observed between the stable and unstable centers, demonstrating the reliability of the signals we observed.} 
\end{figure*}

To rule out the influence of hardware delays or other experimental effects on spectral stability measurements, we selected two color centers with distinct stability characteristics, labeled $OV_a$ and $OV_b$. We performed repeated CW-ODMR measurements switching between the two centers. As shown in \autoref{figS: 交叉}, the results confirm that the observed spectral peak shifts are not caused by measurement errors but truly exists.

\subsection{Long-Time ODMR Stability Tracking}
\label{section:long-tracking}
\begin{figure}[!htbp]
    \includegraphics{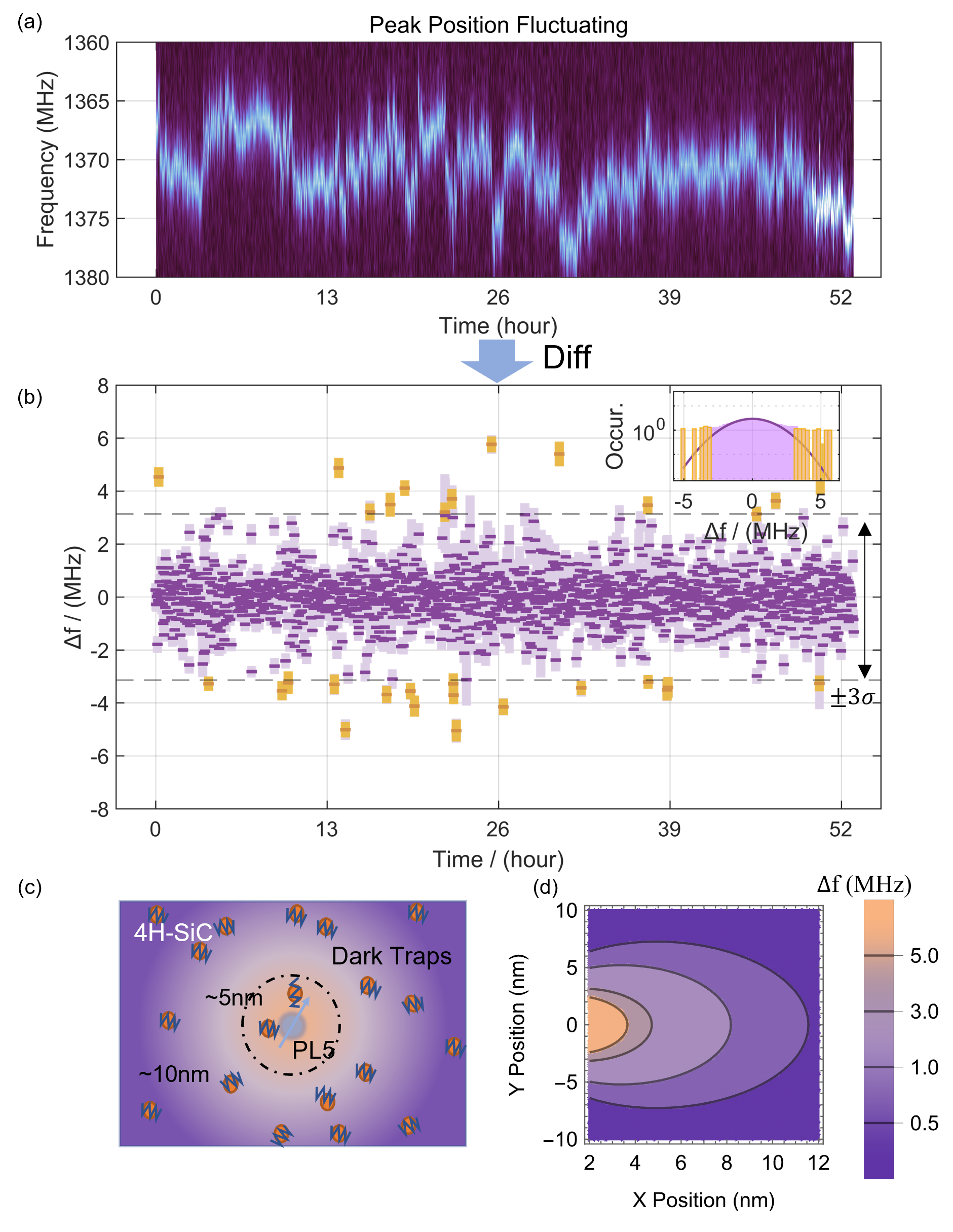}
    \caption{\label{figS: 长时间} (a) CW spectral stability of a single color center monitored over tens of hours. The spectral peak exhibits two-state-like switching behavior together with a long-term gradual drift.
(b) The fitted peak positions were numerically differentiated. Points falling outside the ($±3\sigma$) region are identified as discrete charge-jump events nearby, while the underlying drift is attributed to background noise from the ensemble charge environment. (c) Electrical noise environment around PL5. Under this model, the observed spectral peak shift is contributed by two parts: the electric field drift from distant ensemble defects and the charge release and capture from nearby single trap. After long-duration experiments, the properties of the nearby traps may change and stabilize, leading to an increased contribution from the ensemble drift, and making the two level signal indistinguishable. (d) Relation between the charge location and corresponding stark shift $\Delta_f$, where here we assume the charge is located in the plane of $E_x$.}
\end{figure}

In this case, we performed repeated, long-duration measurements on one of the unstable color centers. Over time, the previously observed two-level signal evolves into a mixed behavior involving both ensemble-level spectral drift and discrete spectral jumps [\autoref{figS: 长时间} (a)]. To analyze this behavior, we apply a differential treatment to the long-term data, as shown in \autoref{figS: 长时间}(b), where the y axis represents the peak shift between adjacent measurements ($\Delta f$). A Gaussian fit reveals large amount of small shift (purple data) together with outlier points beyond $3\sigma$ (orange data). As illustrated in \autoref{figS: 长时间}(c) and (d), we attribute these deviations to a combination of contributions from a broader ensemble of distant charge defects (\textasciitilde 10 nm range) and a smaller number of nearby charges located within \textasciitilde 5 nm of the color center PL5.  \autoref{figS: 长时间}(d) is a numerical result of the relation between a single trap position and the right peak position of PL5 caused by stark shift, where we assume $d_\perp$ =32.5(Hz cm/V) and the trap lies within the same plane as the stress axis of PL5, since the Stark response along the z-direction is weak \cite{2014-ElectricallyMechanicallyTunable-Falk-Phys.Rev.Lett.}. The x-axis is defined along the direction of the local stress field $\vec{E}$ (17.2MHz) of the color center. The calculation result shown in the main text Fig.2(d) was similar  except that we set $d_\parallel$ = 1(Hz cm/V).

\newpage
\subsection{ACF Analysis}
\label{Sec.ACF}
As described in \Cref{section:two color center} and \Cref{section:long-tracking}, high-signal-to-noise ratio (SNR) data resolving discrete single-charge dynamics is rare. In most experimental scenarios, color centers are subject to ensemble-like electronic noise from the background environment, which obscures the distinct switching between discrete states using HMM method and clustering. To overcome this, we choose to employ Autocorrelation Function (ACF) analysis to robustly extract the underlying state fluctuation rate $\Gamma$. 

\begin{figure*}[htbp]
    \includegraphics{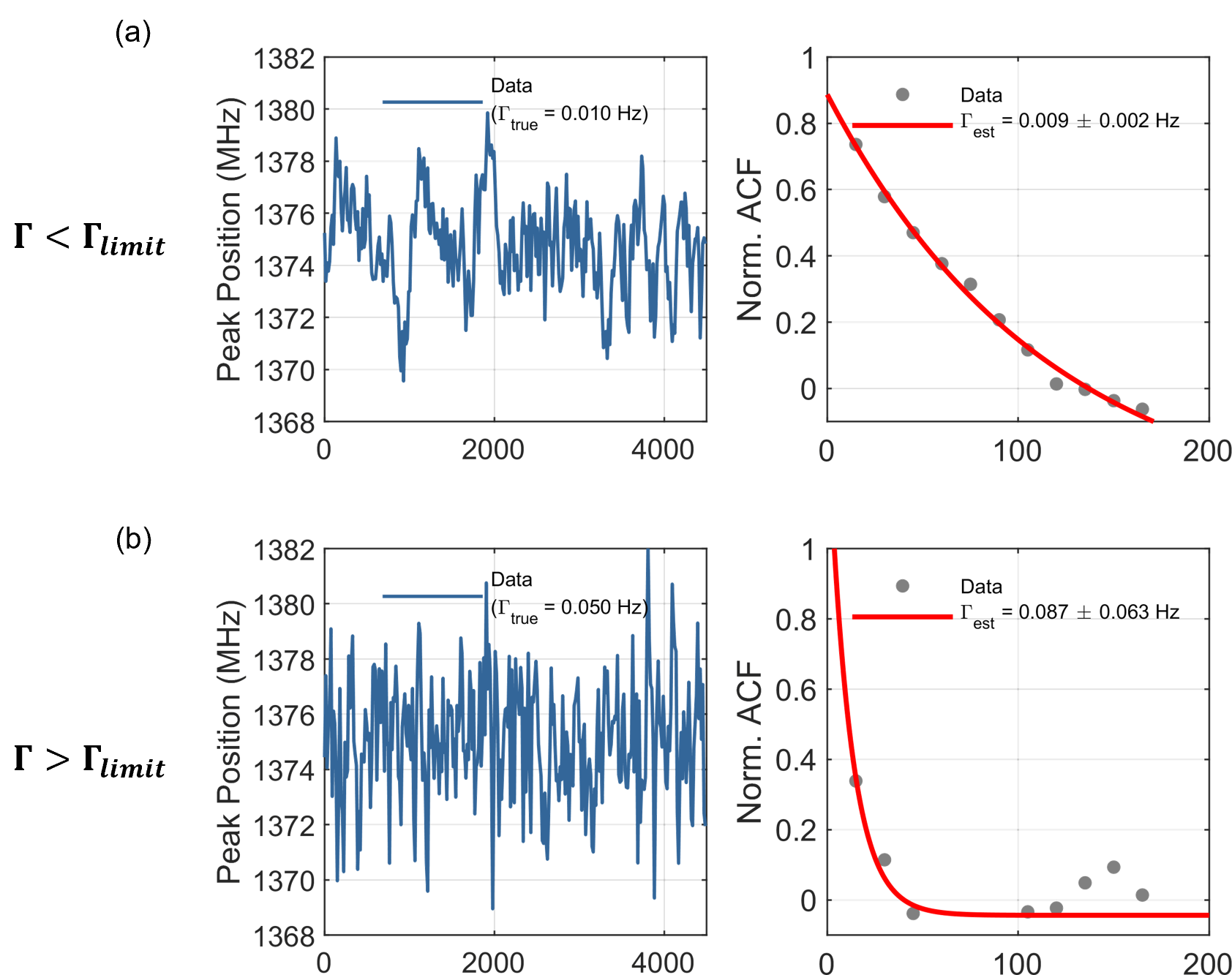}
    \caption{\label{figS: SM_ACF} (a-b) Left panel: A synthetic dataset ($N=300$, $dt=15$ s) generated using an Ornstein--Uhlenbeck process to simulate spectral peak jitter. The model incorporates two different ground-truth transition rate $\Gamma_{true}$ , overlaid with white noise arising from measurement precision and ensemble charge noise. The extracted ACF (gray circles) are derived from the simulation time series. The red solid line represents a nonlinear least-squares fit to the exponential decay model $A \cdot \exp(-\Gamma \tau)+t$. The extracted rate $\Gamma_{est} = 0.009 \pm 0.002$ Hz (1$\sigma$ fit uncertainty) for $\Gamma_{true}=0.01$Hz, which shows excellent agreement with the pre-set physical value, confirming the accuracy of the methodology for experimental data analysis. While for another situation shown in (b) where $\Gamma>\Gamma_{limit}$, the fit is failed with a large error $\Gamma_{est}=0.087 \pm 0.063$ Hz. }
\end{figure*}

To validate the reliability of this approach, we first generated a simulated time series that mimics experimental peak-position jitter under typical noise conditions around 1375 MHz. This allows us to demonstrate the algorithm's capability to recover the physical transition rate even when the raw signal is embedded in significant white noise and baseline drift as shown in \Cref{figS: SM_ACF}. The red solid line represents a nonlinear least-squares fit to the exponential decay model $A \cdot \exp(-\Gamma \tau)+t$. Simulation results demonstrate that exponential fitting can accurately resolve transition rates for dynamics slower than our detection limit. Furthermore, for processes exceeding this limit, we can distinguish between them by analyzing the fitting state and the associated fitting errors.

\begin{figure*}[htbp]
    \includegraphics{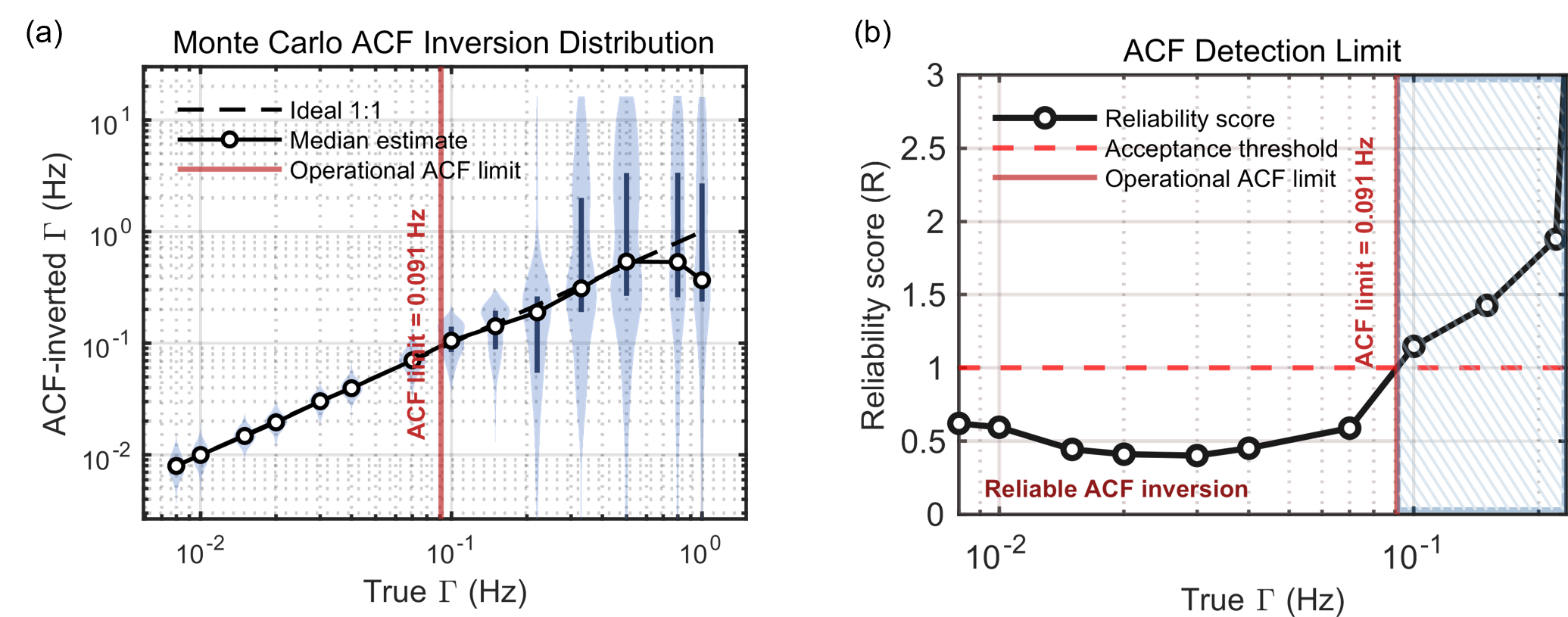}
    \caption{\label{figS: SM_ACF_MC} (a) For each prescribed relaxation rate $\Gamma_{\rm true}$, 300 synthetic peak-position trajectories were generated using an Ornstein--Uhlenbeck process with the same sampling interval, trajectory length, fluctuation amplitude, and white-noise level as in the experiment. 
    The violin plots show the distribution of ACF-extracted rates $\Gamma_{\rm ACF}$, while the black circles indicate the median values. 
    The dashed line denotes the ideal relation $\Gamma_{\rm ACF}=\Gamma_{\rm true}$.  (b) The score $R$ combines the relative median bias and the normalized interquartile range of the Monte Carlo-extracted rates. 
    The ACF inversion is considered reliable for $R<1$, corresponding to a median bias below 30\% and an IQR below 50\% of $\Gamma_{\rm true}$. 
    The red dashed horizontal line indicates the acceptance threshold $R=1$, and the red vertical line gives the resulting operational detection limit of approximately $0.091$ Hz. 
    Beyond this limit, the ACF fit can still return finite rates, but the extracted values are no longer statistically stable.
}
\end{figure*}

To quantify the practical detection limit of the ACF analysis ($\Gamma_{limit}$), we performed Monte Carlo simulations using the same acquisition parameters as in the experiment. The simulated peak-position trajectories were generated from an Ornstein--Uhlenbeck process with a prescribed relaxation rate $\Gamma_{\rm true}$, a stationary fluctuation amplitude of $2.2$ MHz, and an additional white measurement noise of $0.5$ MHz. Each trajectory contained $N=2000$ points sampled every $\Delta t=15$ s. For each value of $\Gamma_{\rm true}$, 300 independent trajectories were generated and analyzed using the same ACF fitting procedure as that used for the experimental data.

The extracted rate distributions broaden as $\Gamma_{\rm true}$ approaches the temporal resolution limit of the ACF method. We therefore define an operational ACF detection limit using both accuracy and precision criteria. Specifically, the ACF inversion is considered reliable only when the relative median bias is below 30\% and the interquartile range (IQR) of the extracted rates is below 50\% of $\Gamma_{\rm true}$. These two conditions are combined into a reliability score R we defined as:
\begin{equation}
\label{S0}
R=\max\left(
\frac{|\mathrm{median}(\Gamma_{\rm ACF})-\Gamma_{\rm true}|/\Gamma_{\rm true}}{0.30},
\frac{\mathrm{IQR}(\Gamma_{\rm ACF})/\Gamma_{\rm true}}{0.50}
\right).
\end{equation}
We define condition $R<1$ as the reliable regime in our experiment. Under the present experimental sampling condition discussed above, this criterion gives an operational ACF detection limit of approximately $0.091$ Hz. Rates above this value may still yield finite fitted values, but the recovered distribution becomes sufficiently broad that the result should no longer be interpreted as a quantitatively reliable ACF inversion.

\subsection{Distribution of Fluctuation Rate across Sensors}
\begin{figure*}[htbp]
    \includegraphics{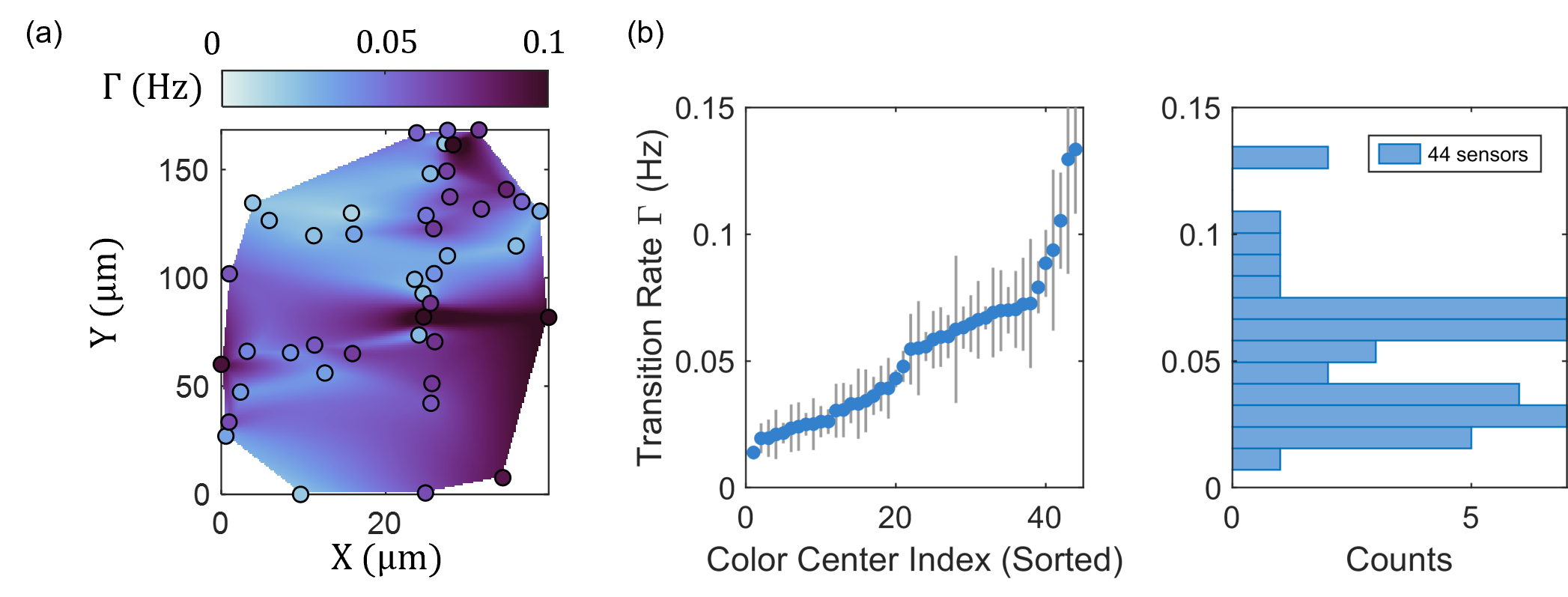}
    \caption{\label{figS: 402ACFImage} Statistical distribution of charge fluctuation rates in Sample 1. (a) Spatial map of the ACF-extracted transition rates ($\Gamma$) for the 44 reliable color center sensors (Fitting $R^2 > 0.7$). (b) Histogram of the corresponding transition rates. The distribution reveals that the local charge environment is dominated by slow dynamic processes, concentrated in the range of 0.02 to 0.1 Hz}
\end{figure*}

We performed ACF analysis on the CW stability data of 64 color centers in Sample 1. By filtering for centers with a fitting coefficient $R^2 > 0.7$, we retained 44 sensors to generate the statistical spatial and rate distributions shown in \Cref{figS: 402ACFImage}. The results indicate that the charge fluctuations in the sample are predominantly associated with slow processes in the frequency range of 0.02 to 0.1 Hz. Faster fluctuations could not be resolved using the ACF method in this experiment.

 Besides, we observed this broad distribution with two apparent characteristic scales around 0.025 and 0.06 Hz. Given the limited statistics, we do not assign a specific microscopic origin. Nevertheless, this result indicates substantial heterogeneity in the local fluctuator environments.

\subsection{Laser Power Dependency of CW Stability}
\begin{figure*}[htbp]
    \includegraphics{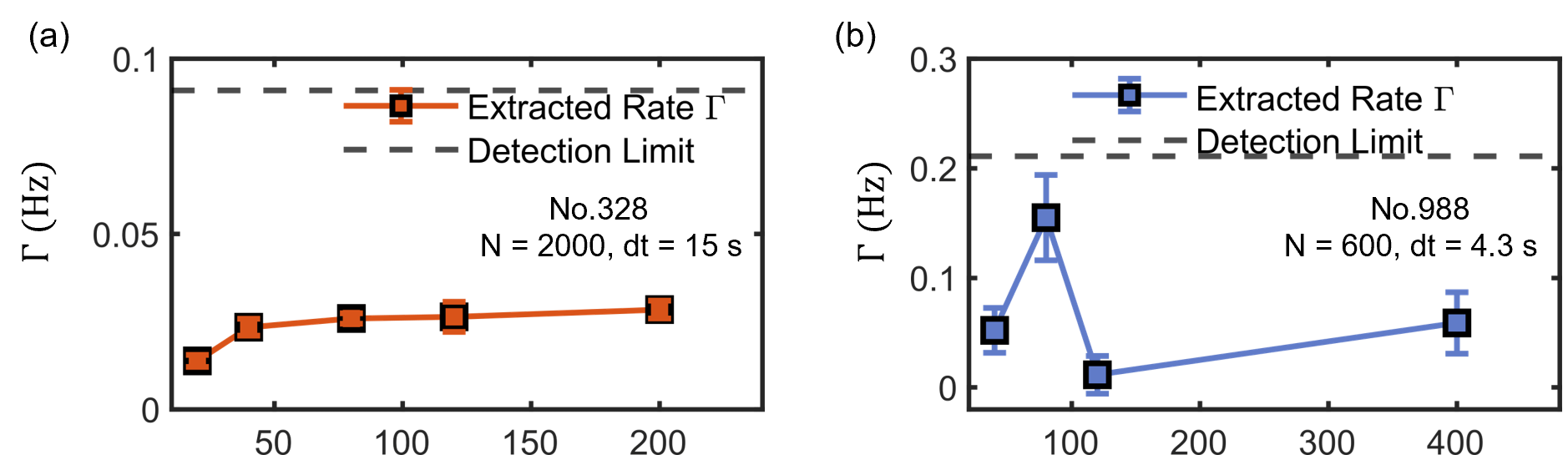}
    \caption{\label{figS: Rate} Power-independent stability of color center dynamics. The ACF-extracted transition rates for different color centers, plotted as a function of optical excitation power. The data indicate that the transition rates exhibit minimal sensitivity or only slightly increase with power variations over an order of magnitude. The SN328 center is consistent with the results presented in the main text. The detection limit is determined according to different sampling parameters N and dt.}
\end{figure*}

To examine whether the observed CW jitter is predominantly governed by optical excitation, we next performed laser-power-dependent measurements on two representative PL5 centers selected from the characteristic rate ranges identified in the statistical ACF distribution above. Using the ACF method described above, we extracted the characteristic time scales of the color centers at different laser powers; the reciprocal value $1/\tau_c$ gives the transition rate $\Gamma$. The two selected PL5 centers correspond to typical local dynamics within the observed distribution \Cref{figS: 402ACFImage}(b), including the No.328 sensor presented in the main text. As shown in \Cref{figS: Rate}, the extracted transition rates show only weak dependence on 905 nm laser power. This behavior is far weaker than expected for a dominant one- or two-photon-driven process. When the optical power changes by one order of magnitude, the increase in the transition rate remains less than threefold. Therefore, while we cannot exclude weak photo-assisted perturbations for rare individual defects, the combined statistical and power-dependent results indicate that the dominant near-DC charge instability observed in Sample 1 is not predominantly governed by optical excitation.

\subsection{Correlation with Magnetic Sensing Range}

\begin{figure*}[htbp]
    \includegraphics{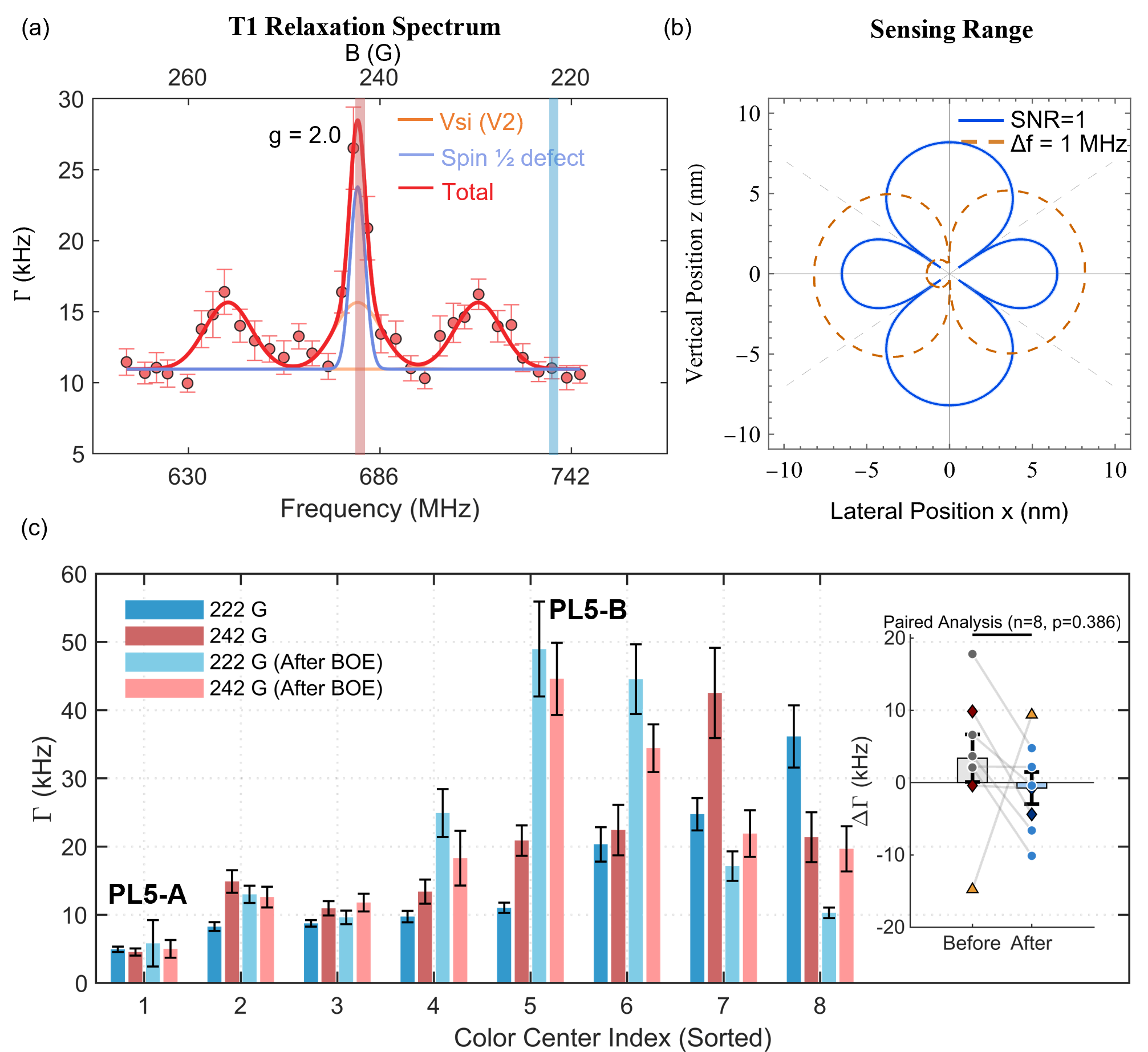}
    \caption{\label{figS: SMT1} (a) EPR Spectrum shown in main text. (b) Sensing range of $T_1$ method. Here, the sensing range boundary is calculated based on the $T_1$ relaxation measurement protocol. Assuming a dark electron spin (S=1/2) with an inhomogeneous linewidth (FWHM) of 6 MHz, the calculated minimum detectable coupling strength is 420 kHz. The blue contour encloses the spatial region where the dark spin can be detected (SNR > 1), yielding a maximum sensing distance of 5$\sim$ 8 nm. This magnetic sensing volume shows a strong spatial overlap with the electrical sensing range derived from Stark shifts \autoref{figS: 长时间}(d), where the dashed line represents the $\Delta f=1MHz$ boundary. (c) EPR comparison across different PL5 sensors before and after BOE treatment. Inset figure shows the g = 2.0 signal strength ($\Delta \Gamma=\Gamma_{242}-\Gamma_{222}$) before/after BOE treatment.}
\end{figure*}

To provide more evidence that the electrical and magnetic noise sources share a common origin, we analyze the sensing range of the $T_1$ relaxation spectroscopy and compare it with the defect distribution derived from the Stark shift analysis (\autoref{figS: 长时间}(d)). The cross-relaxation (flip-flop) interaction between the sensor and a dark spin leads to a resonant reduction in the measured $T_1$ time when their transition frequencies match. Solving the Lindbald equation, the decay rates induced by dark electrons can be expressed as 
\begin{equation}
\label{S1}
    Q=\frac{4\gamma^2\Gamma_D}{\Gamma_D^2+\Delta^2}
\end{equation}
where we assume the dark eletron has similar $T_1$ decay rates with PL5 center ($\sim 10 \mathrm{ kHz}$) and $\Gamma_D$ can be esitimated via the FWHM of spectrum shown in \autoref{figS: SMT1}(a). The sensitivity of our relaxation measurement is given by
\begin{equation}
\label{S2}
    \delta_\Gamma=\frac{2}{\tau\exp(-\tau/T_d)}\frac{1}{\sqrt{N}}\frac{1}{C\sqrt{\alpha}}
\end{equation}

, $\tau$ is the evolution time, $T_d$ is the $T_1$ decay rate of PL5 center and $C, \alpha$ represent the contrast and photon number in a single measurement, $N$ is the experiment cycles in one sec. By calculating the For a system consisting of S=1 and S=1/2, the effective coupling strength is given by the magnetic dipole-dipole interaction:
\begin{equation}
\label{S3}
    \gamma =\frac{1}{2\sqrt{2}} \frac{\mu_0 g_e^2 \mu_B^2}{4\pi \hbar r^3} \left| 1 - 3 \cos^2 \theta \right|
\end{equation}
where $r$ is the distance and $\theta$ is the angle with respect to the quantization axis ($z$-axis). 
\autoref{figS: SMT1}(b) shows the calculated sensing boundary when SNR=1. The maximum sensing radius extends to approximately $5\sim 8$ nm.

Remarkably, this magnetic sensing range aligns well with the electrical charge noise range estimated from the long-term ODMR spectral diffusion shown in \autoref{figS: 长时间}(d). This spatial consistency strongly suggests that the fluctuating charges responsible for the Stark shifts and the dark spins causing $T_1$ relaxation likely originate from the same environment in the vicinity of the sensor. To systematically validate the impact of the BOE treatment on the local spin environment (e.g.,dangling bond), we statistically analyzed the $T_1$ relaxation rates ($\Gamma$) of 8 individual color centers (including PL5-A and PL5-B). As shown in \Cref{figS: SMT1}(c), measurements were performed at two specific magnetic fields: 222 G (off-resonance, representing the broadband high-frequency magnetic background noise) and 242 G (on-resonance, corresponding to the g = 2.0 paramagnetic defects), both before and after BOE treatment. The statistical results yield two key observations. First, the baseline relaxation rates at 222 G (light blue bars) near-universally increase post-BOE across all measured centers. This indicates that the etching process introduces intense, higher-frequency magnetic noise at the newly exposed interface, elevating the overall relaxation background. Second, the relative signal strength of the paramagnetic defects—represented by the difference in relaxation rates between 242 G and 222 G ($\Delta\Gamma=\Gamma_{242}-\Gamma_{222}$)—exhibits an decrease after the BOE treatment for some PL5 (typically, Defect 2, 5 and 7). This suggests a reduction in the density of these specific structural paramagnetic defects upon the removal of the oxide layer. Notably, the variation in $T_1$ and the relative signal strength exhibit distinct differences among individual color centers (inset figure). This implies the complexity of the interfacial spin environment and highlights the heterogeneity of the local nanoscale semiconductor interface experienced by different color centers, which warrants further investigation.

\newpage 
\section{Analysis of Noise Spectroscopy}
\subsection{Dynamical Decoupling and SQ/DQ measurement of PL5 with different ODMR Stability} 

\label{secS: orien-PL5}
\begin{figure*}[htbp]
    \includegraphics[scale=0.8]{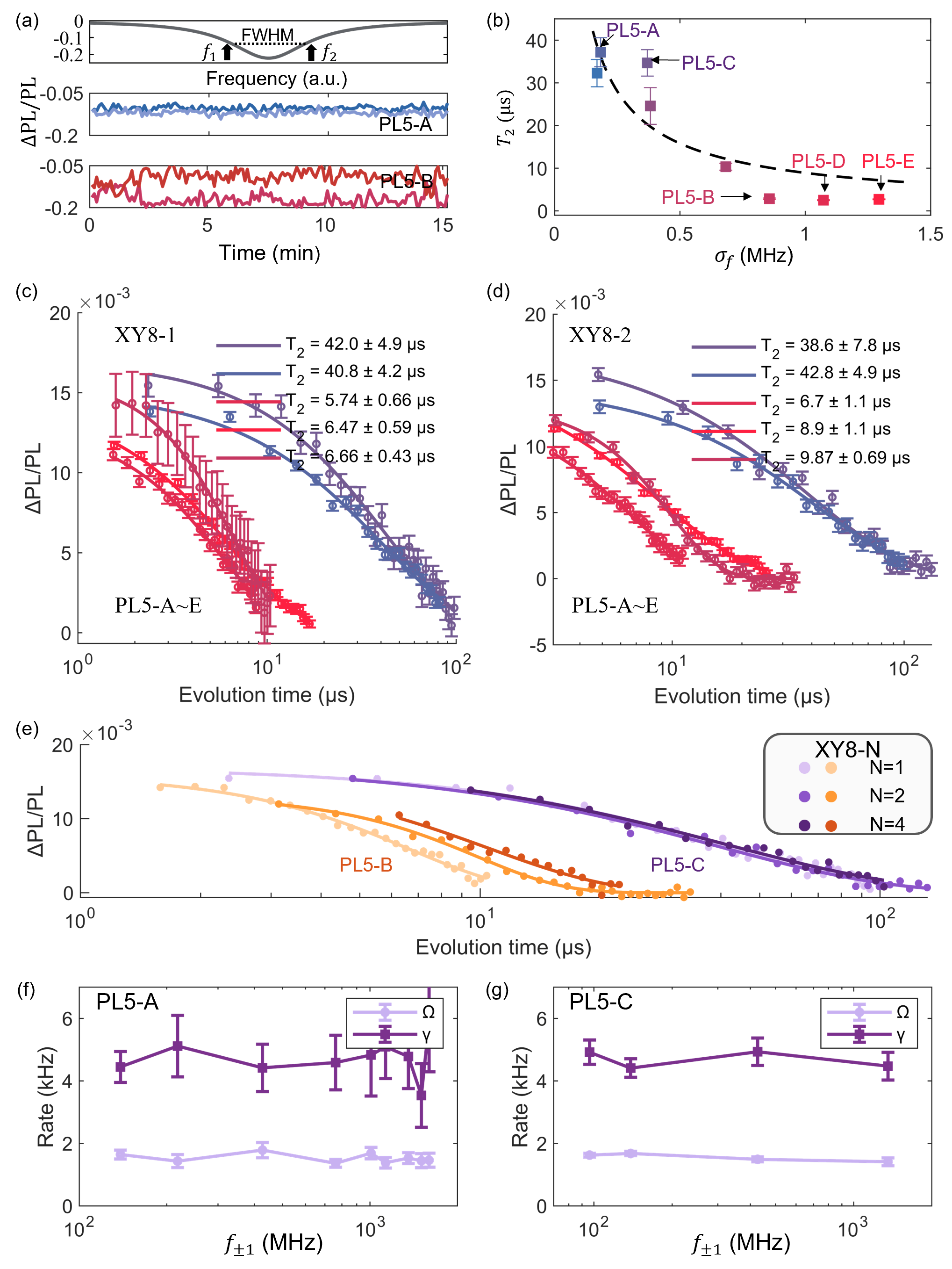}
    \caption{\label{figS: DDSQDQ} (a)Two point measurement method and results of PL5-A and PL5-B. The anti-correlation evolution of the contrast at two fixed frequency position $f_1$ and $f_2$ reveals different stability behavior of peak position. (b) The relation between coherence time (Hahn echo) $T_2$ and their ODMR stability $\sigma_f$ shown in main text. The gray curve serves as a guideline.
(c-d) Comparison of dynamical decoupling results at different pulse orders (XY8-1 and XY8-2) between the two types of centers. (e) Coherence enhancement comparision between PL5-B and PL5-C. (f-g) SQ and DQ rate measurement for two stable quantum sensors PL5-A and PL5-C. The relaxation rates show a consistent behavior, remaining essentially the same ($\gamma/\Omega\approx 2.4$).}
\end{figure*}

We replaced the CW full-spectrum scan with a two-point measurement to achieve higher temporal resolution. We monitored the time evolution of the ODMR contrast at the two half-maximum points of a single spectral peak. Each point pair was acquired in 6 s and we retrieved the peak position from its contrast to calculate the corresponding standard deviation $\sigma_f$. Typical results are shown in \autoref{figS: DDSQDQ}(a), where the two sensors named PL5-A and PL5-B are consistent with the two PL5 used in the main text. We performed dynamical decoupling experiments of varying pulse orders on the color centers presented in the main text. The results and comparation are shown in \autoref{figS: DDSQDQ}(b-e). As shown, for unstable quantum sensors, XY8 and higher-order dynamical-decoupling sequences can prolong the coherence time, yet they cannot elevate it to the level attained by stable ones.
For color centers with stable spectral peaks, named PL5-A and PL5-C, we performed and compared their SQ and DQ relaxation rates [\autoref{figS: DDSQDQ}(f-g)]. The results show that within the measured frequency range, the ratio of relaxation rates for DQ and SQ transitions remain essentially the same ($\gamma/\Omega\approx 2.4, \Omega\equiv \Omega_+=\Omega_-$).
It should be noticed that for PL5-B, $(\gamma/\Omega)_{@1GHz}\approx 0.9$, much lower than PL5-A/C, which may indicate a broad magnetic noise background.   

Besides, since our measurements on PL5 centers involve environments with significant low-frequency (near-DC) noise, we provide a detailed analysis here to verify the integrity of the dynamical decoupling (DD) data in the presence of frequency drifts. 

Firstly, the XY8 sequence is intrinsically robust against static detuning for arbitrary initial states \cite{1990-NewCompensatedCarrPurcell-Gullion-JournalofMagneticResonance1969}. In this pulse configuration, pulse imperfections and off-resonance effects only influence the DD fidelity to the second order \cite{2012-ComparisonDynamicalDecoupling-Wang-Phys.Rev.B}. Regarding the DD filter function, the resonance peak undergoes only negligible modulation when the detuning is small compared to the Rabi frequency of the control pulses \cite{2019-NonvanishingEffectDetuning-Lang-Phys.Rev.A}. In our experiments on PL5-B, the Rabi frequency is approximately $6.25 \text{ MHz}$, while the detuning—estimated from the ODMR stability $\sigma_f$—is on the order of $0.7 \text{ MHz}$. These parameters fall within the regime discussed in \cite{2019-NonvanishingEffectDetuning-Lang-Phys.Rev.A}, where the spectral weight of the filter function remains largely conserved. This ensures that the noise spectroscopy $\mathcal{S}(f)$ reconstructed in the main text remains accurate despite slow detuning drifts. Secondly, the frequency drift occurs on a timescale of seconds, which is several orders of magnitude slower than the total duration of a single XY8 sequence ($\sim \mu\text{s}$). Consequently, each individual measurement cycle can be treated under a quasi-static approximation, where the detuning is constant and effectively compensated since XY8 pulse configuration is robust to quasi-static detuning.

\subsection{Comparison of Noise Spectroscopy of PL5-A and B with BOE Treatment}
\begin{figure*}[htbp]
    \includegraphics[scale=0.92]{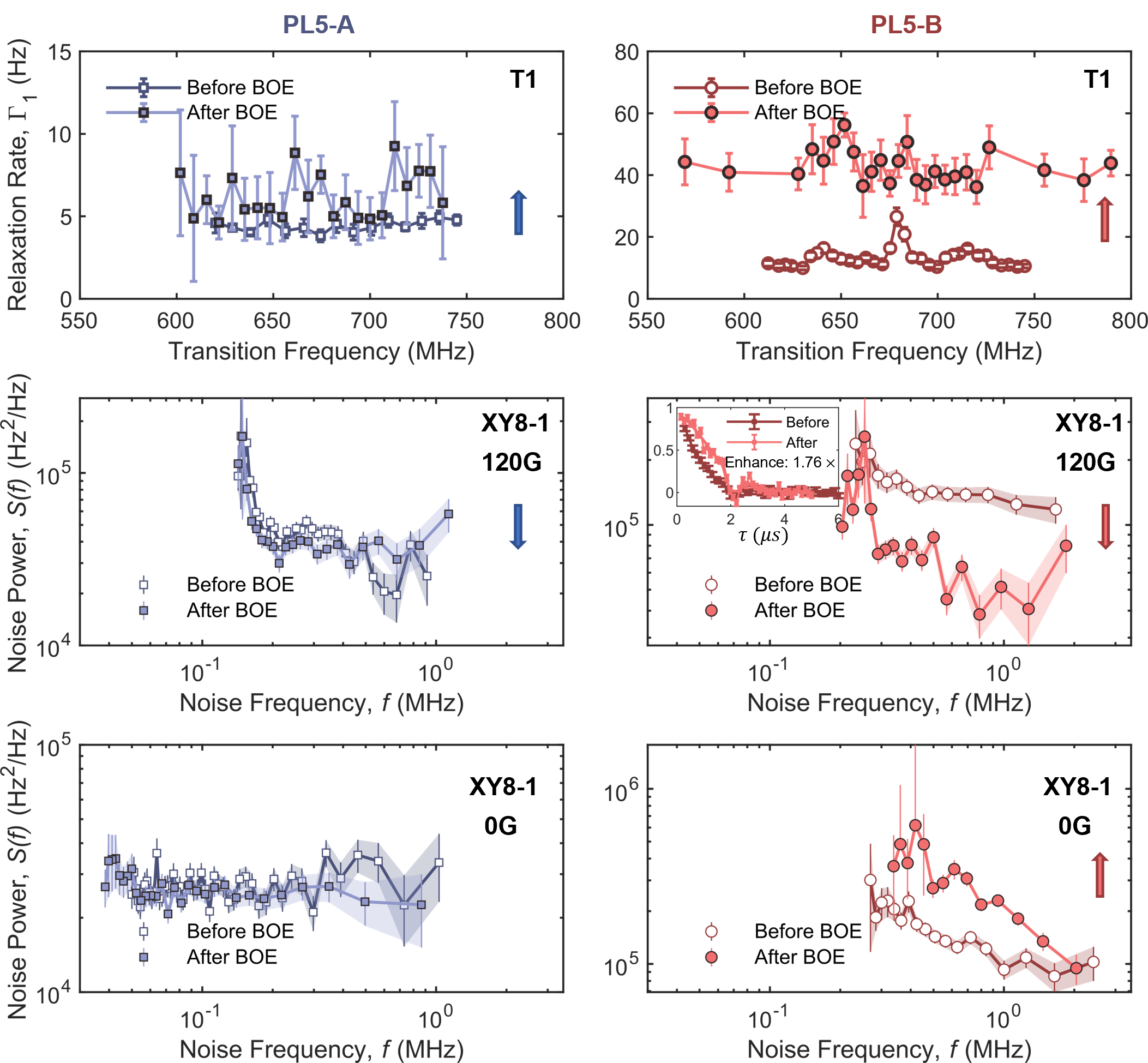}
    \caption{\label{figS: PL5AB_BOE} Frequency-dependent reconfiguration of the noise spectrum following BOE treatment. The removal of the native oxide layer exerts a contrasting impact on the local electromagnetic noise across different frequency bands. $T_1$ relaxation measurements reveal a significant enhancement of magnetic noise in the GHz regime post-treatment. Conversely, dynamical decoupling under an axial magnetic field demonstrates a pronounced suppression of magnetic noise in the MHz regime. These results highlight the trade-off of surface treatment, where the elimination of the slow-to-intermediate oxide spin bath comes at the cost of increased high-frequency fluctuations from the bare surface.}
\end{figure*}

The samples were first cleaned by ultrasonic cleaning procedure in organic solvents and then immersed in a buffered oxide etchant (BOE, 7:1) for 30 s to remove the native oxide layer. Following removal, the samples were rapidly rinsed in deionized water, dried with a nitrogen gun, and quickly immersed in oil to prevent prolonged exposure to the ambient environment. We subsequently relocated the PL5-A and PL5-B sensors to perform comparative measurements. As shown in the \Cref{figS: PL5AB_BOE}, both centers exhibit consistent noise responses following the treatment as desrribed in main text. The coherence time of the PL5-B center under an axial magnetic field was significantly prolonged, enabling the resolution of previously obscured coupled nuclear spin signals that were previously inaccessible due to the limited coherence lifetime. We noted that the influence of the reconstructed noise landscape is slightly less pronounced in PL5-A than in PL5-B, likely due to a difference in their respective depths below the surface.

\newpage
\subsection{Cross-check of T1 Relaxometry by DEER Measurement}
\begin{figure*}[htbp]
    \includegraphics{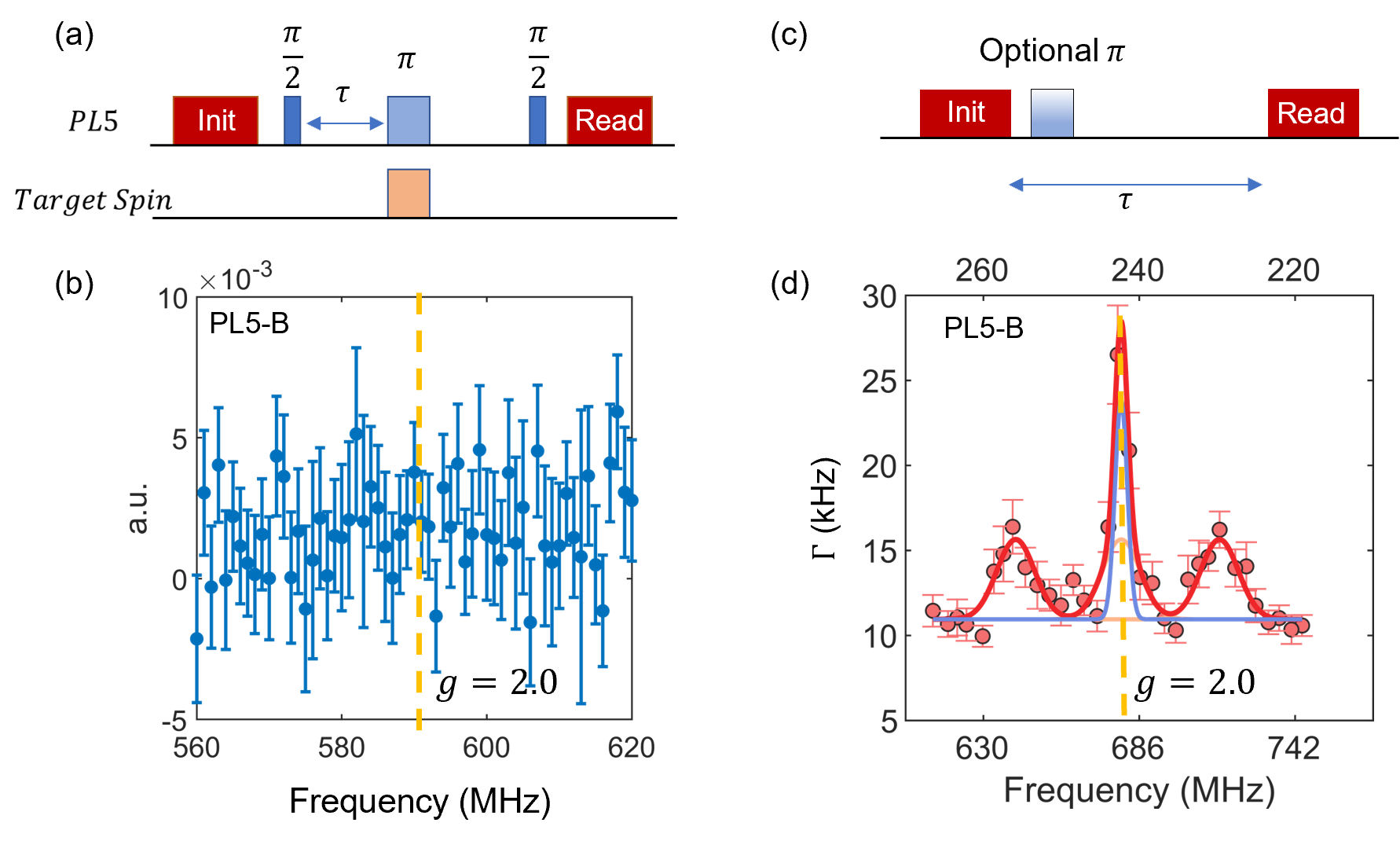}
    \caption{\label{figS: DEER} (a) DEER pulse sequence used to probe the local target spin near PL5-B.
    (b) DEER spectrum measured on PL5-B. No measurable DEER response is observed under the present pulse conditions. The DEER evolution time is $\tau = 1~\mu$s.
    (c) $T_1$-relaxometry sequence.
    (d) Corresponding $T_1$-EPR spectrum of the same PL5-B center, showing a clear resonance fingerprint. The dashed line marks the expected $g = 2.0$ transition.}
\end{figure*}

To further examine the local spin-noise fingerprint identified by $T_1$ relaxometry, we performed a DEER measurement on the same PL5-B center at the magnetic field where the $T_1$-EPR spectrum shows the V2-related resonance. As shown in \Cref{figS: DEER}, no measurable DEER response is observed for this local target under the present pulse conditions, whereas $T_1$ relaxometry clearly resolves the corresponding resonance fingerprint. This comparison does not imply that DEER is generally unable to detect local spin environments. Rather, it reflects the complementary detection channels of the two methods: DEER relies on coherent driving of the target transition, while $T_1$ relaxometry detects resonant incoherent fluctuations through the sensor longitudinal relaxation channel. Therefore, for the specific local environment examined here, $T_1$ relaxometry provides access to local dark-spin noise that is not readily resolved by coherent double-resonance detection under our experimental conditions.
\newpage
\section{Comparison with Conventional Ensemble-Averaged Characterizations}
To comprehensively evaluate the macroscopic electrical and defect environments of the SiC wafers and to benchmark our PL5 nanoscale sensor against conventional ensemble techniques, standard continuous-wave electron paramagnetic resonance (CW-EPR) measurements, Deep-level transient spectroscopy (DLTS) and  electrical property test (I-V and C-V) were performed as discussed below. 

\subsection{EPR Measurement}
\begin{figure*}[htbp]
    \includegraphics{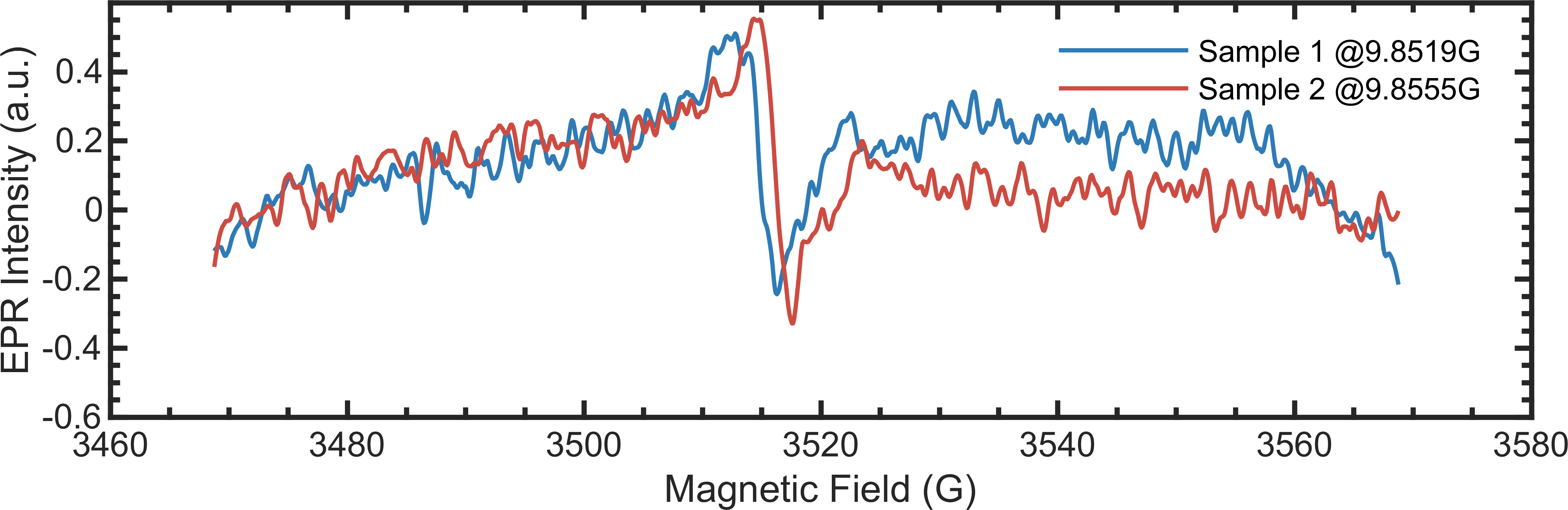}
    \caption{\label{figS: SM_EPRCompare}  Room-temperature continuous-wave EPR spectra of the SiC wafers. The EPR spectrum was acquired using a Ciqtek EPR200M X-Band spectrometer. Both Sample 1 and Sample 2 exhibit a similar resonance feature near $g = 2.0027$. The fluctuating baseline is primarily attributed to microwave cavity loading and the skin effect induced by the conductive SiC substrate. }
\end{figure*}

To evaluate the macroscopic defect environments of the SiC wafers, standard continuous-wave electron paramagnetic resonance (CW-EPR) measurements were performed at room temperature. As shown in Figure \Cref{figS: SM_EPRCompare}, both Sample 1 and Sample 2 exhibit a similar volume-averaged resonance centered at $g \approx 2.0027$, a nearly isotropic signal typically attributed to carbon dangling bonds or near-surface amorphous carbon-rich clusters \cite{2000-CharacterizationParamagneticDefect-Macfarlane-J.Appl.Phys.,SiC-book}. The undulating baseline observed in the spectra might be attributed to microwave cavity Q-factor degradation and skin depth limitations caused by the highly doped SiC substrates.

Notably, our EPR results do not exhibit distinct characteristic peaks originating from intrinsic vacancy defects. This observation is physically reasonable. In traditional EPR studies, detecting vacancy-related centers typically requires subjecting the samples to high-dose electron irradiation or high-concentration ion implantation to generate defect densities that exceed the macroscopic detection limit \cite{2002-ContinuouswavePulsedEPR-Mizuochi-Phys.Rev.B,SiC-book,2001-CarbonVacancyrelatedDefect-Son-Phys.Rev.B,2012-NegativeSystemCarbon-Son-Phys.Rev.Lett.}. Furthermore, specific intrinsic defects, such as the carbon vacancy ($V_C$), exhibit negative-U properties, whose paramagnetic signals can only be clearly resolved at cryogenic temperatures combined with specific optical excitation \cite{2012-NegativeSystemCarbon-Son-Phys.Rev.Lett.}. The absence of these signals in our room-temperature measurements highlights the fundamental sensitivity limitations of conventional defect characterization methods. 

First, while ensemble ESR provides averaged information over a large number of defects, single PL5 centers reveal pronounced sensor-to-sensor variations within the same wafer, highlighting the heterogeneous nature of the local defect environment.

More crucially, for investigating the specific properties within the epitaxial layer and the interface between epi and $\ce{SiO_2}$ layer — which is our primary region of interest, conventional EPR is inherently inadequate. Because standard EPR captures a volume-averaged signal from the entire bulk crystal, the weak response of the thin epitaxial layer (tens of micrometers) is completely obscured by the overwhelming background of the thick substrate (hundreds of micrometers)\cite{SiC-book}. Consequently, accurately isolating the defect signature of the epilayer using traditional EPR would necessitate the destructive physical removal of the substrate \cite{2012-NegativeSystemCarbon-Son-Phys.Rev.Lett.}. These inherent bottlenecks of macroscopic characterization demonstrate the unique advantages of nanoscale sensing platform.

\subsection{I-V / C-V characterization}

To further characterize the macroscopic electrical properties of the wafers, we utilized two $5 \times 5 \text{ mm}^2$ pieces cut from the exact same Wafer 1 and Wafer 2 used in the main text, without any ion implantation or thermal annealing, named Sample 1-0 and Sample 2-0. 

\subsubsection{Device Fabrication} After identical standard wafer cleaning procedures, a Ni/Au (100 nm / 100 nm) metal stack was deposited on the backside of both wafers, followed by Rapid Thermal Annealing (RTA) at $950^{\circ}\text{C}$ to form the ohmic contacts simultaneously. Subsequently, a top Ti/Au (10 nm / 150 nm) metal stack was deposited to define the $500\mu m \times 500 \mu m$ square Schottky contacts, as shown in \Cref{figS: SM_CVIV}(a-b).

\subsubsection{Electrical Characterization} 
\begin{figure*}[htbp]
    \includegraphics[scale=0.75]{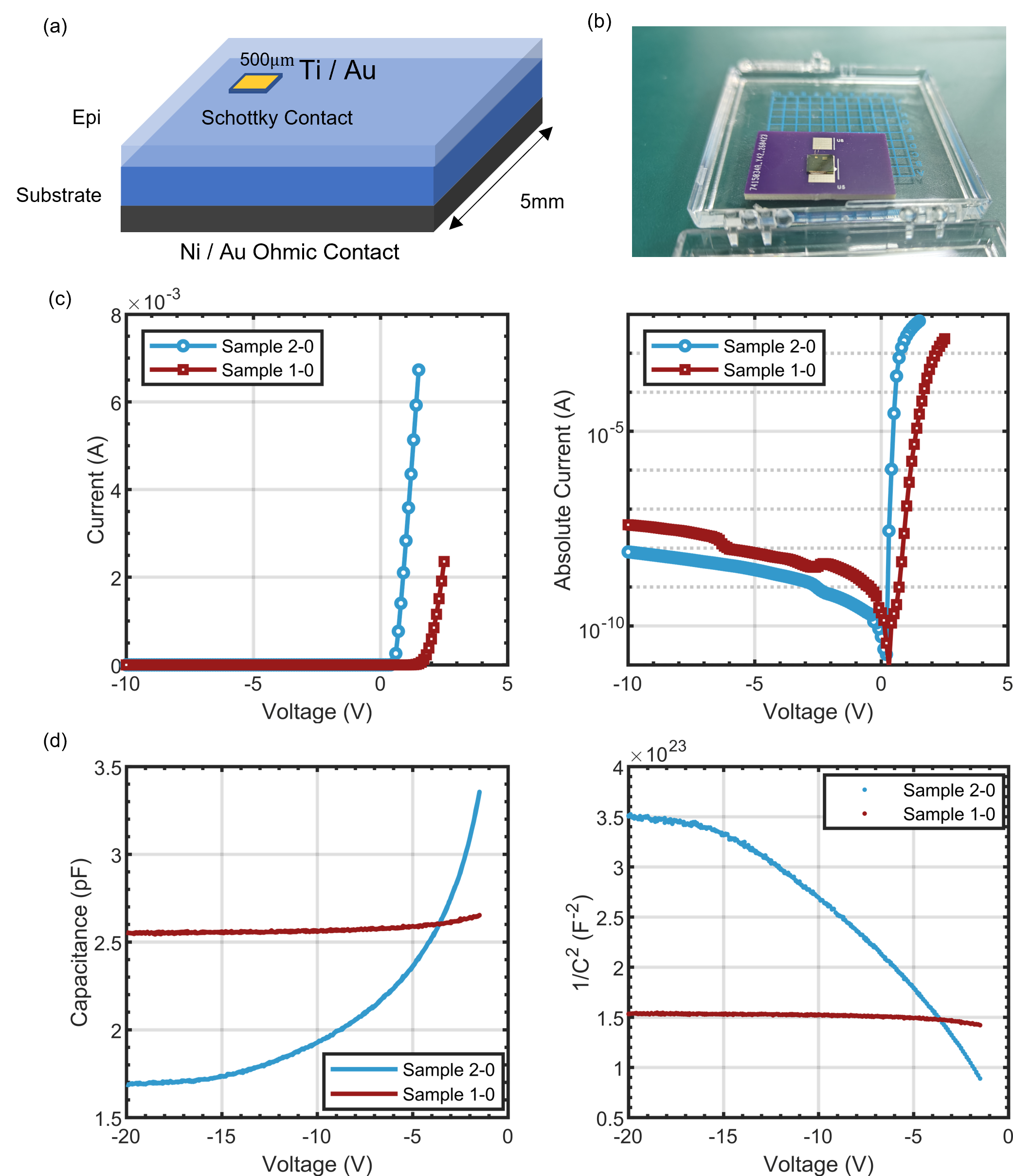}
    \caption{\label{figS: SM_CVIV} (a) Schematic of the 4H-SiC Schottky diode structure with top Ti/Au (10 nm / 150 nm, $500 \times 500 \mu m^2$) Schottky contacts and bottom Ni/Au (100 nm / 100 nm) ohmic contacts. (b) Photograph of the fabricated $5 \times 5\text{ mm}^2$ 4H-SiC chips placed in a chip tray. (c) Linear and semi-logarithmic current-voltage ($I-V$) characteristics comparing Sample 1-0 and Sample 2-0  at room temperature. Sample 1-0 exhibits a significantly delayed turn-on voltage, higher ideality factor, and increased reverse leakage current. (d) Capacitance-voltage ($C-V$) and corresponding $1/C^2-V$ characteristics at 1 MHz under reverse bias. Sample 2-0 shows typical depletion layer expansion, whereas Sample 1-0 exhibits severe capacitance pinning, indicating a high density of interface states or a thick interfacial layer hindering the modulation of the depletion region. }
\end{figure*}

Despite the identical fabrication process, the conventional $I-V$ measurements  reveal a stark contrast between the two samples. Sample 2-0 demonstrates near-ideal Schottky diode behavior with a sharp forward turn-on and extremely low reverse leakage current. In contrast, Sample 1-0 exhibits severe non-ideal characteristics, including a significantly delayed turn-on voltage (shifted by nearly 1.0 V), a severely degraded ideality factor, and elevated reverse leakage current. These strongly suggest the presence of a substantial parasitic series resistance or an interfacial barrier layer. The 1 MHz $C-V$ and $1/C^2-V$ profiles (\Cref{figS: SM_CVIV}, Keithley 4200A) provide deeper physical insight into this degradation. For Sample 2-0, the capacitance decreases non-linearly with increasing reverse bias, following the relationship for abrupt junctions. Conversely, the $C-V$ curve of Sample 1-0 exhibits a severe ``pinning" effect—the capacitance remains almost entirely flat ($\sim $2.5 pF) regardless of the applied reverse bias up to -20 V. The corresponding $1/C^2$ slope approaches zero, physically implying that the depletion region in the 4H-SiC epitaxial layer fails to expand under an external electric field. 

\begin{figure*}[htbp]
    \includegraphics[scale=1]{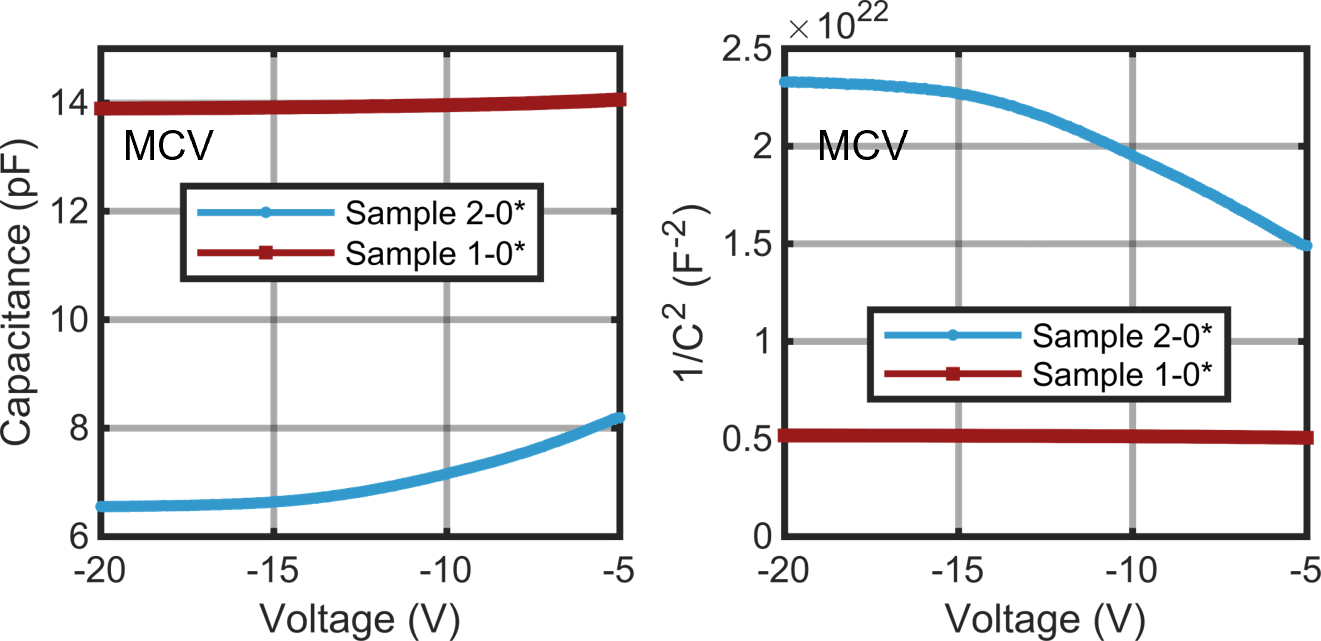}
    \caption{\label{figS: SM_MCV}  Mercury-probe capacitance-voltage (MCV) characteristics. $C-V$ and $1/C^2-V$ profiles of bare slices from Wafer 1 (red) and Wafer 2 (blue). Consistent with the fabricated diodes (\Cref{figS: SM_CVIV}), the sample from Wafer 1 (corresponding to Sample 1-0*) exhibits severe capacitance pinning, confirming that the severe degradation originates from macroscopic surface deterioration on the wafer itself rather than the fabrication process.}
\end{figure*}

It is worth noting that to ensure the reliability of this difference, we performed mercury-probe capacitance-voltage (MCV) measurements—a conventional technique for wafer inspection—on other bare slices from Wafer 1 and Wafer 2 without metal electrodes \Cref{figS: SM_MCV}. The obtained results were consistent, confirming that the observed surface differences between Wafer 1 and Wafer 2 represent a macroscopic and deterministic variation. This indicates extremely high-density surface states or a thick, non-ideal interfacial layer that completely screens the applied voltage in Wafer 1.
\newpage

\newpage
\subsection{DLTS Measurement}
\begin{figure*}[htbp]
    \includegraphics[scale=1]{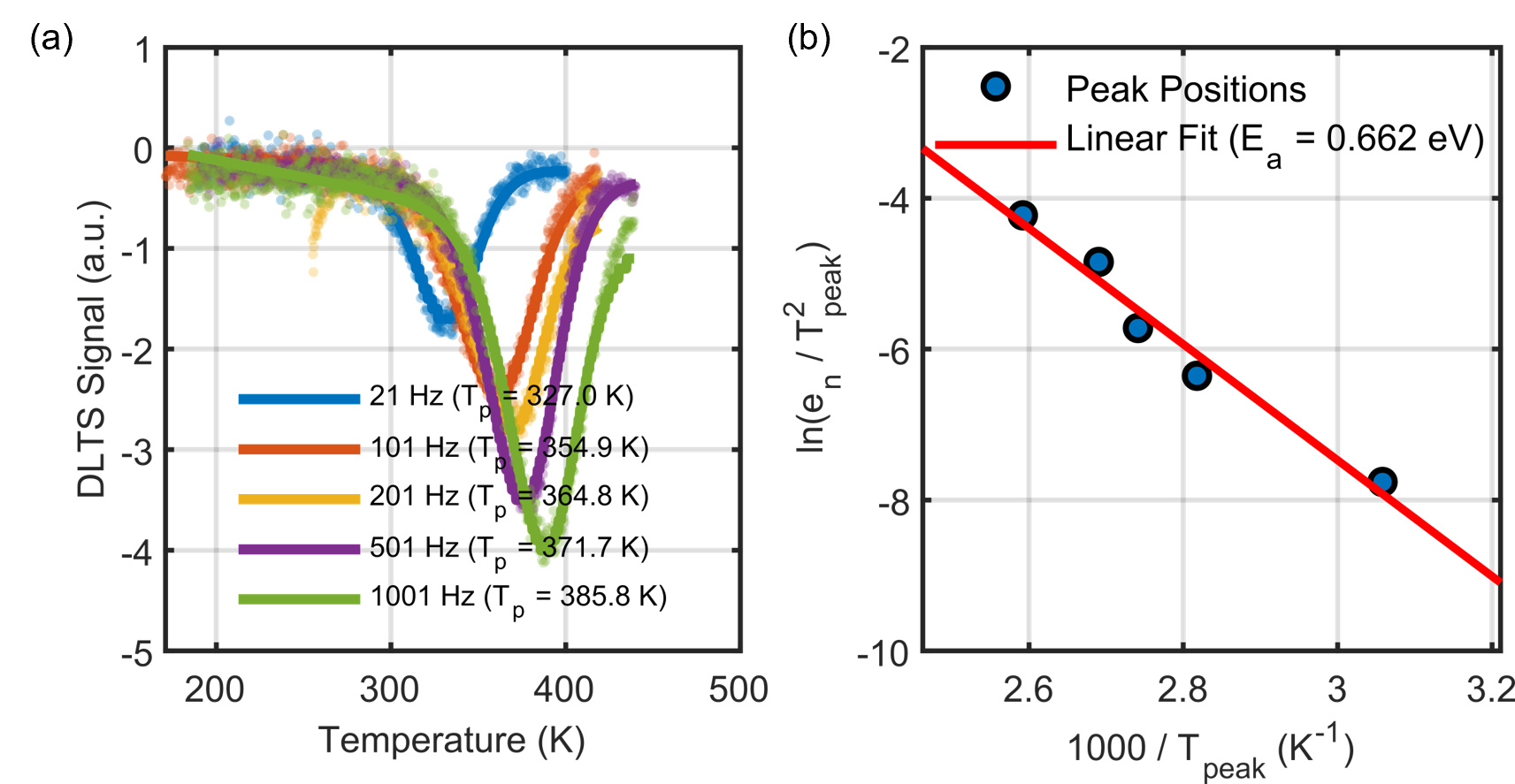}
    \caption{\label{figS: SM_DLTS} Characterization of deep-level defects via DLTS on sample 2-0. Deep-level transient spectroscopy (DLTS) scans performed on Wafer 2 using a Semilab DLS-83D spectrometer. Note that measurements on Wafer 1 yielded no signals due to poor interface quality obscuring the capacitance variation. The excitation pulse width was maintained at 20 $\mu$s throughout the scans. Arrhenius plots constructed by shifting the lock-in frequencies reveal a prominent trap level whose activation energy closely matches the fingerprints of carbon vacancies ($Z_{1/2}$).}
\end{figure*}

Deep-level transient spectroscopy (DLTS) is a powerful technique for characterizing deep-level defects in semiconductor epitaxial layers by monitoring the capacitance transients resulting from the thermal emission of charge carriers following a voltage pulse. Here, we applied DLTS to analyze the defect distribution in our SiC samples. However, for Wafer 1, the high density of interfacial defects leads to significant pinning of the Fermi level and parasitic capacitance effects, rendering the standard C–V based DLTS measurements totally ineffective. Consequently, we present DLTS results exclusively for Wafer 2 (\Cref{figS: SM_DLTS}). Through Arrhenius analysis of the emission rates, we identified characteristic signatures of carbon vacancy ($V_C (Z_{1/2})$) defects within the epitaxial layer \cite{doi:https://doi.org/10.1002/9781118313534.ch5,1997-DeepLevelDefects-Hemmingsson-J.Appl.Phys.}. It is important to note that standard C-V-based DLTS depends on a well-defined capacitance response and is therefore less effective for Wafer 1, where the capacitance is strongly pinned. The PL5 quantum-sensing approach, by contrast, directly probes the local interfacial noise environment through an embedded near-surface sensor, revealing nanoscale charge-noise dynamics and nearby spin-related defect fingerprints at room temperature. This provides information complementary to conventional transport-based characterization, particularly in cases where macroscopic capacitance-based methods lose diagnostic sensitivity.

